\documentclass[12pt]{article}
\usepackage{latexsym}
\usepackage{amsthm}
\usepackage{amsmath}
\usepackage[dvips]{graphicx}
\usepackage{amssymb}
\usepackage{multirow}

\begin{document}
\baselineskip 0.6cm

\def\simgt{\mathrel{\lower2.5pt\vbox{\lineskip=0pt\baselineskip=0pt
           \hbox{$>$}\hbox{$\sim$}}}}
\def\simlt{\mathrel{\lower2.5pt\vbox{\lineskip=0pt\baselineskip=0pt
           \hbox{$<$}\hbox{$\sim$}}}}

\begin{titlepage}

\begin{flushright}
UCB-PTH-10/17\\
\end{flushright}

\vskip 2.0cm

\begin{center}

{\LARGE \bf 
Origins of Hidden Sector Dark Matter I: \\ \vspace{0.25cm} Cosmology
}

\vskip 1.0cm

{\large Clifford Cheung$^{1,2}$, Gilly Elor$^{1,2}$, Lawrence J. Hall$^{1,2,3}$ and Piyush Kumar$^{1,2,4}$}

\vskip 0.4cm

$^{1}${\it Berkeley Center for Theoretical Physics, Department of Physics, \\
     University of California, Berkeley, CA 94720} \\
$^{2}${\it Theoretical Physics Group, Lawrence Berkeley National Laboratory,
     Berkeley, CA 94720} \\

$^{3}${\it Institute for the Physics and Mathematics of the Universe, \\ University of Tokyo, Kashiwa 277-8568, Japan}\\


$^{4}${\it Department of Physics,  Columbia University, New York, NY 10027}

\abstract{

We present a systematic cosmological study of a universe in which the visible sector is coupled, albeit very weakly, to a hidden sector comprised of its own set of particles and interactions.  Assuming that dark matter (DM) resides in the hidden sector and is charged under a stabilizing symmetry shared by both sectors, we determine all possible origins of weak-scale DM allowed within this broad framework.  We show that DM can arise only through a handful of mechanisms, lending particular focus to Freeze-Out and Decay and Freeze-In, as well as their variations involving late time re-annihilations of DM and DM particle anti-particle asymmetries.  Much like standard Freeze-Out, where the abundance of DM depends only on the annihilation cross-section of the DM particle, these mechanisms depend only on a very small subset of physical parameters, many of which may be measured directly at the LHC.  In particular, we show that each DM production mechanism is associated with a distinctive window in lifetimes and cross-sections for particles which may be produced in the near future.  We evaluate prospects for employing the LHC to definitively reconstruct the origin of DM in a companion paper.

}

\end{center}
\end{titlepage}

\def\simgt{\mathrel{\lower2.5pt\vbox{\lineskip=0pt\baselineskip=0pt
           \hbox{$>$}\hbox{$\sim$}}}}
\def\simlt{\mathrel{\lower2.5pt\vbox{\lineskip=0pt\baselineskip=0pt
           \hbox{$<$}\hbox{$\sim$}}}}

\renewcommand{\l}{\langle}
\renewcommand{\r}{\rangle}
\newcommand{\be}{\begin{eqnarray}}
\newcommand{\ee}{\end{eqnarray}}

\newcommand{\dd}[2]{\frac{\partial #1}{\partial #2}}
\newcommand{\NN}{\mathcal{N}}
\newcommand{\LL}{\mathcal{L}}
\newcommand{\MM}{\mathcal{M}}
\newcommand{\ZZ}{\mathcal{Z}}
\newcommand{\WW}{\mathcal{W}}

\newcommand{\FO}{{\rm FO}}
\newcommand{\FI}{{\rm FI}}
\newcommand{\TFO}{T_{\rm FO}}
\newcommand{\TFOp}{T_{\textrm{FO}'}}
\newcommand{\YFO}{Y_{\rm FO}}
\newcommand{\zFO}{z_{\rm FO}}
\newcommand{\FOp}{\textrm{FO}'}

\newcommand{\sv}{\langle \sigma v \rangle}

\newcommand{\MPl}{M_{\rm Pl}}

\section{Introduction}

As our understanding of theoretical high-energy physics has evolved, top-down considerations have motivated the exploration of ``parallel sectors" comprised of their own particles and interactions but ``hidden" from us (the visible sector) due to the weakness of the couplings connecting hidden  and visible sector particles. In addition to providing new avenues for model-building, this broad framework opens up a range of exciting possibilities for the origin of Dark Matter (DM), which is the focus of this work. Understanding the origin of DM and its interactions within this framework is very important because experimental observations have only measured the gravitational effects of DM, leaving a large number of logical possibilities. 


Assuming that the visible sector and other possible hidden sectors are initially in a state of thermal equilibrium,  what are the possible production mechanisms for DM? If DM shares sizeable interactions with visible sector particles, then thermal equilibrium will be efficiently maintained until Freeze-Out (FO) renders a thermal relic abundance of DM via the standard WIMP paradigm \cite{Kolb:1988aj}.  Alternatively, it may be that DM couples extremely weakly to the visible sector and to itself, as is the case for so-called superWIMPs \cite{Feng:2003uy,Feng:2004zu} and FIMPs \cite{Hall:2009bx,Asaka:2005cn}.  A third and final possibility is that DM is very weakly coupled to the visible sector, but has substantial couplings to a hidden sector to which it is thermally equilibrated.  In general, this hidden sector will contain its own set of particles and interactions and will have a temperature \emph{different} from that of the visible sector\footnote{If the temperature of the two sectors are the same, the sectors have equilibrated implying that there is only one sector.}.   The purpose of the present work is to systematically identify and characterize all possible origins of DM which might arise in this enormous class of theories.  

We will assume throughout that DM is stable due to a symmetry shared by the visible and hidden sectors.  Moreover, let us denote the lightest visible and hidden sector particles charged under this symmetry by $X$ and $X'$, which have masses $m$ and $m'$ taken to be broadly of order the weak scale such that $m> m'$.  By definition, $X'$ is the DM particle.  We also assume the existence of a weak coupling which bridges the visible and hidden sector and mediates the decay
\be
X \rightarrow X' + \ldots,
\ee
where the ellipses denote what are typically visible decay products.  

Remarkably, the cosmological evolution of this setup is entirely fixed by only a handful of parameters.  This is analogous to standard single sector FO, where the DM abundance is solely determined by the DM annihilation cross-section.  Here we find that DM relic abundance is fixed by following set of parameters in general: 
\be
\{ m, m', \langle \sigma v\rangle ,  \langle \sigma v \rangle',  \xi , \tau, \epsilon\},
\label{eq:params}
\ee
where $\langle \sigma v\rangle$ and $\langle \sigma v \rangle'$ are the thermally averaged annihilation cross-sections for $X$ and $X'$, respectively, $\xi$ is the ratio of the visible and hidden sector temperatures, $\tau$ is the lifetime of $X$, and $\epsilon$ is a measure of the CP-phase in $X$ decays. In particular cases, the relic abundance depends on only a subset of the above parameters, as will be shown below.

We have evolved the cosmological history of the visible and hidden sectors over the parameter space defined in Eq.~(\ref{eq:params}) in order to systematically identify all possible origins of hidden sector DM.  Of course, the simplest possibility is that DM undergoes hidden sector Freeze-Out (FO$'$), yielding a thermal relic abundance.  This has been considered in many hidden sector models, and was studied systematically in \cite{Feng:2008mu}.  On the other hand, the remaining possibilities for the origin of DM fall into two very broad categories:  
\begin{itemize}
\item {\bf Freeze-Out and Decay (FO\&D).}  $X$ undergoes FO and then decays out of equilibrium, yielding an abundance of $X'$.  As we will see later, the final abundance of $X'$ goes as
\be
\Omega &\propto& 
\frac{m'}{m\langle \sigma v\rangle}.
\label{eq:FODeq}
\ee
\item {\bf Freeze-In (FI). } $X$ decays while still in thermal equilibrium with the visible sector, yielding an abundance of $X'$. As we will see later, the final abundance of $X'$ goes as
\be
\Omega & \propto&  
\frac{m'}{m^2 \tau}.
\label{eq:FIeq}
\ee
\end{itemize}
Within the categories of FO\&D and FI exist a number of distinct variations.  For example, if FO\&D or FI happen to produce an abundance of $X'$ particles exceeding a particular critical value, then the $X'$ particles will promptly undergo an era of ``re-annihilation.''  During this time the $X'$ particles will efficiently annihilate within a Hubble time despite the fact that $X'$ is no longer thermally equilibrated with the hidden sector.  Because the final DM abundance changes accordingly, we refer to this mechanism of DM production as FO\&D$_{\rm r}$ and FI$_{\rm r}$.   Another variation arises if $X$ decays are CP-violating, in which case FO\&D and FI may produce an abundance of DM endowed with a particle anti-particle asymmetry. Such an effect is possible because although the visible and hidden sectors are separately in thermal equilibrium, they are not in equilibrium with each other. We denote these asymmetric modes of DM production by Asymmetric Freeze-Out and Decay (FO\&D$_{\rm a}$) and Asymmetric Freeze-In (FI$_{\rm a}$).  Note that these mechanisms are entirely distinct from the framework of Asymmetric DM \cite{Kaplan:2009ag}, in which the DM particle anti-particle asymmetry is inherited from an already existent baryon asymmetry.

\begin{figure}[t]
  \center{
\includegraphics[scale=1]{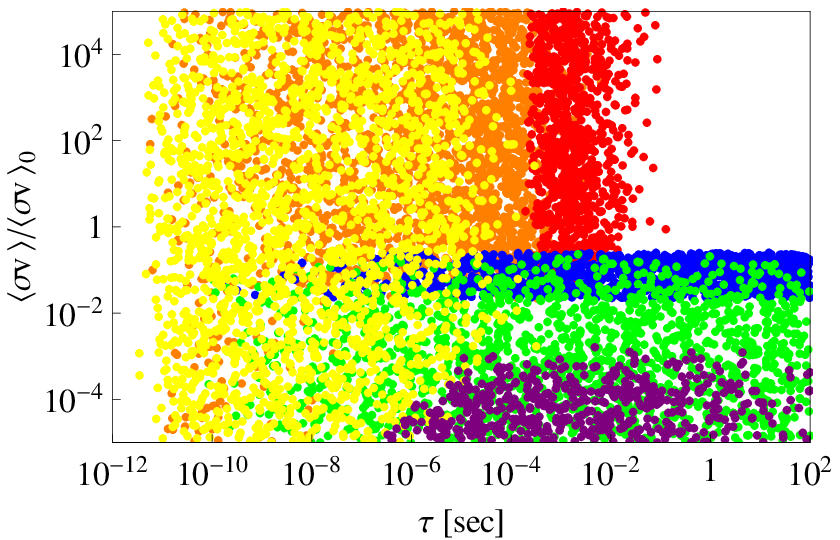}
\includegraphics[scale=1]{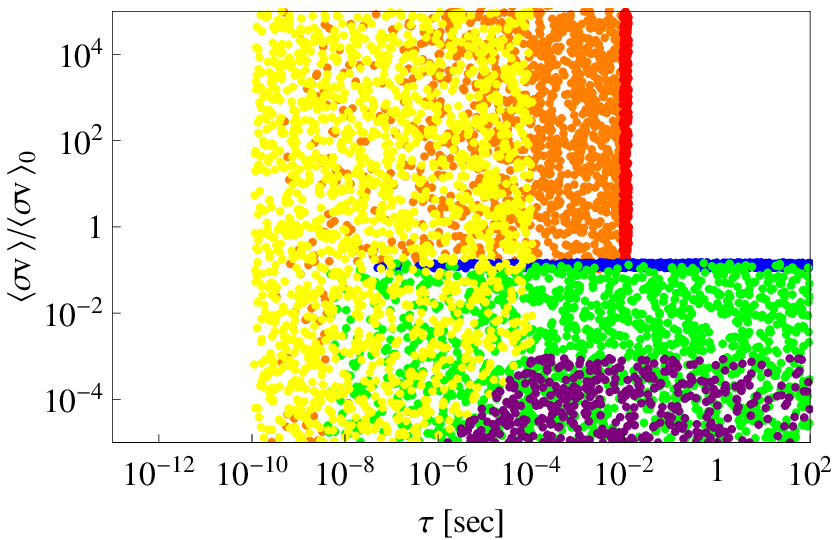}
}
\caption{\footnotesize{Hidden sector DM can originate via a handful of production mechanisms, each corresponding to a distinctive window in the $\tau - \langle \sigma v\rangle$ plane.  Aside from hidden sector FO$'$, these mechanisms are \{FO\&D, FO\&D$_{\rm r}$, FO\&D$_{\rm a}$, FI, FI$_{\rm r}$, FI$_{\rm a}$\}, denoted by \{blue, green, purple, red, orange, yellow\}.  
Each point corresponds to $\Omega h^2 =0.11$, where we have scanned over a very inclusive parameter space defined by $10^{-5} < \langle \sigma v\rangle /\langle \sigma v\rangle_0,\langle \sigma v\rangle '/\langle \sigma v \rangle_0 < 10^5$, $10^{-3} < \xi < 10^{-1}$, $10^{-8} < \epsilon < 10^{-3} $, where  $\langle \sigma v\rangle_0 = 3\times 10^{-26} \textrm{ cm}^3/\textrm{s}$.  In the left panel, the masses have been scanned over a broad region $10\textrm{ GeV} < m < 1\textrm{ TeV}$ and $1/20 < m'/m < 1/2$, while in the right panel, the masses have been fixed to a narrow region $m=100 \textrm{ GeV}$ and $1/4 < m'/m < 1/3$.   }}
\label{fig:introplot}
\end{figure}

Crucially, as seen in Eqs.~\ref{eq:FODeq} and \ref{eq:FIeq}, each of these DM production mechanisms maps to a rather distinctive window in the parameter space spanned by $\tau$ and $\langle \sigma v\rangle $---and where {\it all} other parameters, $m$, $m'$, $\langle \sigma v\rangle ' $, $\xi$, and $\epsilon$, are scanned over an inclusive range of values.  This is remarkable because $\tau$ and $\langle \sigma v\rangle $ can, in principle, be measured at the LHC---after all, they are attributes of $X$, which is a visible sector field!   For example, see the left panel of Figure \ref{fig:introplot}, where each point corresponds to $\Omega h^2 = 0.11$, and each color denotes the dominant mechanism of DM production at that particular point in parameter space.  Even though all parameters but $\tau$ and $\langle \sigma v\rangle$ have been scanned over a generous range, one sees that FO\&D corresponds to a narrow band in $\langle \sigma  v\rangle $ while FI corresponds to a narrow band in $\tau$. 

On the other hand, it is also very likely that $m$, and perhaps even $m'$, might be measured at colliders, for instance if the visible decay products of $X \rightarrow X' + \ldots$ can be used to kinematically reconstruct the event.  In the event that $m$ and $m'$ are indeed measured, the boundaries between DM production mechanisms in the $\tau - \langle \sigma v\rangle$ plane become even more distinct, as shown in the right panel of Figure \ref{fig:introplot}.

Because each production mechanism lies in a distinctive region in the $\tau -  \langle \sigma v\rangle$ plane, we are left with the tantalizing possiblity that the origin of DM might be successfully reconstructed at the LHC even in this much broader framework compared to that of standard single sector FO. The purpose of the present work, however, is to establish a comprehensive understanding of the structures depicted in Figure \ref{fig:introplot}, leaving a more detailed collider study to a companion paper \cite{colliderpaper}.

The outline of this paper is as follows. In Section \ref{sec:twosec} we present a broad overview of two-sector cosmology.  We begin with an analysis of our setup in a decoupled limit in which the visible and hidden sectors couple only through gravitational interactions.  We then introduce portal interactions, and present a detailed discussion of the FO\&D and FI mechanisms of DM production.  Afterwards, we go on to discuss the effects of ``re-annihilation'', followed by an analysis of the thermal properties of the coupled two-sector system.  In Section \ref{cpd}, we present a series of cosmological phase diagrams depicting the dominant production mechanisms for DM as a function of the parameter space.  We go on to discuss how the boundaries in these phase diagrams change with various parameters.    In Section \ref{sec:dma} we present a discussion of DM production from particle anti-particle asymmetries, and we conclude in Section \ref{sec:conclusions}.

\section{Overview of Two-Sector Cosmology}
\label{sec:twosec}

Our setup is comprised of a visible and hidden sector, each with sizeable self-interactions which serve to maintain thermal equilibrium in each sector at temperatures $T$ and $T'$, respectively. We assume that these sectors couple to one other only through portal interactions which are extremely feeble, so these temperatures are not equal, i.e.~$T \neq T'$.   To begin, we limit the present discussion as well as that of Sections \ref{sec:visFO} and \ref{sec:hidFO} to the case in which the visible and hidden sectors are entirely decoupled but for gravitational effects.  In Section \ref{subsection:Portal} and onwards we introduce portal interactions connecting the visible and hidden sectors and study the significant impact of these couplings on the cosmological history.

Throughout, we assume that the visible and hidden sectors enjoy a symmetry, discrete or continuous, that keeps DM cosmologically stable.  The lightest visible sector particle charged under this stabilizing symmetry is denoted by $X$, and likewise in the hidden sector, $X'$, which we take to be lighter than $X$.  In the limit in which the visible and hidden sectors are decoupled, $X$ and $X'$ are, of course, simultaneously stable.  However, as portal interactions are switched on, $X$ becomes unstable and decays with a width $\Gamma$ into particles which ultimately yield an $X'$ in the final state.  We will study the cosmological evolution of the number densities $n$ and $n'$ of $X$ and $X'$, which obey the coupled Boltzmann equations
\be\label{eq:Bn}
\frac{d}{dt} n + 3 H n &=& - (n^2 - n_{\rm eq}^2) \sv - \Gamma n \\
\label{eq:Bn'}
\frac{d}{dt} n' + 3 H n' &=& - (n'^2 - n'^2_{\rm eq}) \sv' + \Gamma n,
\ee
where $n_{\rm eq}$ and $n'_{\rm eq}$ are the thermal equilibrium abundances and it is understood that here $\Gamma$ is thermally averaged and we work in a regime in which the effects of the corresponding inverse decays are negligible.   Here we take the thermally averaged annihilation cross-sections, $\sv$ and $\sv'$, to be independent of temperature.  A primary aim of this paper is to study the most general cosmological evolution which follows from these equations, subject only to the requirement that $\sv$ and $\sv'$ are large enough that both $X$ and $X'$ undergo freeze-out.  Note that we take the masses of $X$ and $X'$, $m$ and $m'$,  both to be broadly of order the weak scale.   

The relative size of $T$ and $T'$ can have a drastic impact on the cosmological history.  To see why this is so, let us define the ratio of temperatures to be 
\be 
\xi & \equiv&\frac{ T'}{T}.
\label{eq:xi}
\ee  
If we assume the standard picture of slow-roll inflation, then the inflaton can, in principle, couple with different strengths to the visible and hidden sectors.  As a consequence, the decay of the inflaton reheats each sector to a different temperature, corresponding to an initial condition for $\xi$ given by $\xi_{\rm inf} = T'_{\rm inf}/T_{\rm inf}$, the ratio of temperatures in each sector immediately after the decay of the inflaton, which we take to be less than 1.  Naively $\xi = \xi_{\rm inf}$ for all time.  However, interactions between the sectors can change $\xi$ from $\xi_{\rm inf}$. For example, scatterings between the sectors which are generically dominated in the UV, can increase the high temperature value of $\xi$ to $\xi_{\rm UV}$, which is taken to be a free parameter. In addition, there can be IR contributions to $\xi$ as well. These contributions are discussed in more detail in Section \ref{subsection:equil}. 

Even ignoring interactions between the two sectors, $\xi$ actually varies as a function of temperature due to the separate conservation of the co-moving entropies, $S=g_{*S}T^3$ and $S'=g'_{*S}T'^3$, in each sector.  Specifically, this implies that $\xi$ varies as a function of temperature to the extent to which the numbers of degrees of freedom in the visible and hidden sectors vary with temperature:
\be
\xi(T) \, \propto \, \left( \frac{g_{*S}(T)}{g'_{*S}(T)} \right)^{\frac{1}{3}},
\label{eq:xiT}
\ee
where $g_{*S}(T)$ and $g'_{*S}(T)$ are the number of relativistic degrees of freedom in the visible and hidden sectors, respectively, when the visible sector is at a temperature $T$.  A change in $g_{*S}(T)$ or $g'_{*S}(T)$ by an order of magnitude only affects $\xi$ at the level of a factor of two; hence, when comparing DM production from processes at different temperatures, this effect may be justifiably ignored.  

In general, the energy density in the hidden sector affects the expansion rate of the universe during BBN, which places an important, albeit weak constraint on $\xi$.  In particular, any hidden sector particles which are relativistic at BBN contribute an effective number of extra neutrino species
\be 
 \Delta N_\nu \, = \,  \frac{4}{7} \,\, g_*'(T_{\rm BBN}) \,\,   \xi(T_{\rm BBN})^4.
\ee 
The present bound from experiment is $ \Delta N_\nu < 1.4$ \cite{Amsler:2008zzb}, which is surprisingly mild: that is, for $g_*'(T_{\rm BBN})=100$, this is satisfied by taking $\xi(T_{\rm BBN})=1/3$.  Furthermore, according to Eq.~(\ref{eq:xiT}), at higher temperatures $\xi$ can be close to unity even if $g'_{*}>100$ \cite{Feng:2008mu}.

\subsection{Visible Sector Freeze-Out (FO)}
\label{sec:visFO}

In the early universe, visible sector particles reside in a thermal bath at temperature $T$ with abundances fixed accordingly by equilibrium thermodynamics. As $T$ drops below the mass of $X$, $m$, the number density of $X$ particles, $n$, remains in thermal equilibrium and undergoes the usual Boltzmann suppression.  The $X$ particles undergo FO as the rate of annihilations, $n \sv$, drops below the expansion rate $H$, which occurs at a temperature $\TFO \simeq m/x_{\rm FO}$.  The parameter $x_{\rm FO}$ depends only logarithmically on $\sv$, and for roughly weak scale cross-sections $x_{\rm FO} \approx 20-25$.  Defining the yield, $Y=n/s$ where $s$ is the entropy of the visible sector, we obtain the familiar expression for the yield at FO, 
\be
\YFO &\simeq& \frac{3}{2\pi} \sqrt{\frac{5}{2}}\frac{\sqrt{g_*}}{g_{*S}}\frac{1}{  \MPl \sv   } \frac{1}{T_\FO}.
\label{eq:YFO}
\ee
In the decoupled limit, $X$ is stable and will account for the totality of DM in the universe if $m \YFO \simeq 4\times 10^{-10} \textrm{ GeV}$.  This corresponds to a critical cross-section of $\sv_0 \simeq 3\times 10^{-26} \textrm{ cm}^3/\textrm{s}$.

Because we have thus far assumed that the visible and hidden sectors only interact gravitationally, the only effect of the hidden sector on FO in the visible sector is through its contribution to the energy density of the universe.  However, this effect is tiny and can be accounted for in Eq.~(\ref{eq:YFO}) by increasing $g_{*}$ by a factor
\be
1 + \frac{7}{43}  \, \Delta N_\nu \, \left( \frac{g_{*}(\TFO)}{g_{*}(T_{\rm BBN})} \right)^{\frac{1}{3}}
\left( \frac{g'_{*}(T_{\rm BBN})}{g'_{*}(\TFO)} \right)^{\frac{1}{3}},
\ee
where we have ignored the difference between $g_*$ and $g_{*S}$.
For $\TFO < M_W$ this increases $\YFO$ by at most 17\% for $ \Delta N_\nu =1$.  Thus the BBN constraint implies that the standard relation between the DM abundance and the DM annihilation cross-section is preserved to a good accuracy.




\subsection{Hidden Sector Freeze-Out (FO$'$)}
\label{sec:hidFO}

\begin{figure}[h!]\label{lspfo}
\center{\includegraphics[scale=0.85]{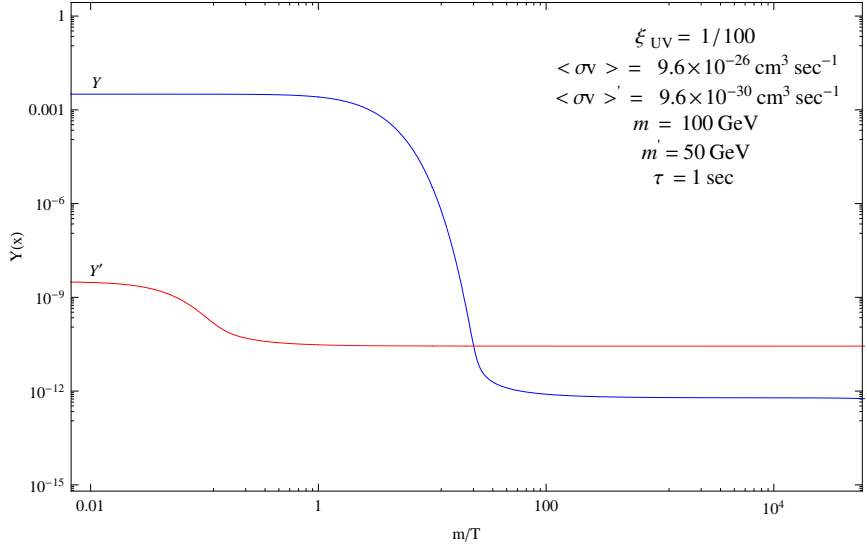}}
\caption{\footnotesize{$X$ and $X'$ yields as a function of $m/T$ for a particular choice of parameters.  The visible and hidden sectors separately undergo FO and FO$'$, respectively.   At high temperatures $Y'/Y \sim \xi_{\rm UV}^3 = 10^{-6}$.  Since $m'$ is not much less than $m$, a cool hidden sector, $\xi \ll 1$, implies that FO$'$ occurs before FO.  A cool hidden sector tends to make $Y'_{\FO'} < Y_\FO$, but this is more than compensated by having $\sv' \ll \sv$. Using the quoted parameters in Eq. (\ref{eq:YFOprime}) gives $Y'_{\FO'} \simeq 200 \, Y_\FO$. }}
\label{fig:FO}
\end{figure}

As the temperature $T'$ of the hidden sector falls below $m'$, the number density of $X'$ particles, $n'$, tracks the equilibrium distribution and becomes exponentially suppressed.  Ultimately, the $X'$ particles undergo FO$'$ once the hidden sector drops to a temperature $T'_{\textrm{FO}'} = m' /x'_{\rm FO'}$, just as $X$ undergoes FO at $T_\FO = m/x_{\rm FO}$.    Here the prime on $T'$ indicates that the temperature is that measured in the hidden sector, while the prime on FO$'$ indicates that this temperature is being evaluated at the time of FO$'$, {\it not} FO.  Thus, the ratio of visible sector temperatures at FO$'$ compared to FO is 
\be 
\frac{\TFOp}{T_\FO} &=& \frac{1}{\xi_{\rm FO'}}\frac{m'}{m},
\label{eq:TFOprime}
\ee  
where we took $x'_{\rm FO'} \simeq x_{\rm FO}$.
Consider the parametric scaling of the above expression. Reducing $m'/m$ tends to shift FO$'$ to a later time than FO, while reducing 
$\xi_{\rm FO'}$ tends to do precisely the opposite.  While in principle either ordering is possible, we focus on a scenario in which FO$'$ occurs before FO, as this leads to a richer set of cosmological histories once portal interactions between the sectors are included.

First, consider the case $\langle \sigma v \rangle'= \langle \sigma v \rangle$.  Because $n'$ at FO$'$ is fixed by $H$ at $T_{\FO'}$ and $T_{\FO'}> T_{\FO}$, there is naively a greater number of $X'$ particles yielded by FO$'$ than $X$ particles yielded by FO.  However, because FO$'$ occurs earlier, there is also a commensurately greater amount of entropy dilution, so the total $X'$ yield  actually turns out to be less than the $X$ yield.  To see this, let us define $Y' = n'/s$ to be the $X'$ yield normalized to the {\it visible sector entropy}, which will be useful for comparing with the $X$ yield.  We find that for arbitrary $\langle \sigma v \rangle'/ \langle \sigma v \rangle$
\be
\frac{Y'_{\FO'}}{Y_\FO}  &\simeq&  \frac{\TFO}{ \TFOp}\frac{\sv}{\sv'} \, \simeq  \, \xi_{\rm FO'}\frac{m}{m'}\frac{\sv}{\sv'}.
\label{eq:YFOprime}
\ee
Hence, the yield of $X'$ from FO$'$ is subdominant to the yield of $X$ from FO as long as the hidden sector is sufficiently cool or if its annihilations are sufficiently strong. See Figure \ref{fig:FO} for a plot of the evolution of $X$ and $X'$ abundances as a function of $x=m/T$, for a choice of parameters making the contribution from FO$'$ greater than that from FO.  

\subsection{The Portal}
\label{subsection:Portal}
Until now, we have not considered the effect of direct, albeit tiny couplings which might directly connect the visible and hidden sectors.  Consider a portal operator $\mathcal O$ which connects $X$ and $X'$, thereby mediating the decay
\be\label{eq:XtoX'}
X &\rightarrow & X' + \ldots ,
\ee
where the ellipses denote what is typically visible SM particles.  
For the moment, let us ignore the particulars of $\cal O$ and attempt to characterize the gross features of the cosmological history as a function of the $X$ lifetime, $\tau = 1/\Gamma$.  As the lifetime is taken from cosmological scales to microscopic scales, the cosmology typically transitions through four broadly defined scenarios\footnote{This is only a rough sketch; a more precise understanding of the various possibilities is given in Section \ref{cpd}.}:
\begin{itemize} 
\item {\bf Multi-Component Dark Matter.}  $X$ is so long lived that it is stable over cosmological time scales.  Thus $X$ and $X'$ comprise the DM of the universe.
\item {\bf Freeze-Out and Decay.}  $X$ decays late, after leaving thermal equilibrium, yielding a contribution to the $X'$ abundance.   
\item {\bf Freeze-In.} $X$ decays fast enough that it produces a substantial $X'$ abundance from decays occurring while $X$ is still in thermal equilibrium.
\item  {\bf Thermalized at Weak Scale.} $X$ decays so quickly that the visible and hidden sectors are actually in thermal equilibrium at the weak scale.  From the point of view of cosmology, the visible and hidden sectors are a single sector.
\end{itemize} 
While the first category is certainly a logical possibility, it has been well explored in the literature and is hard to test experimentally since the DM abundance depends on $\sv'$, so we will ignore it.  Moreover, we will not consider the last category because we are specifically interested in cosmological scenarios in which the visible and hidden sectors are not thermally equilibriated at the weak scale.  Thus, our discussion will center on the FO\&D and FI phases of the two-sector cosmology.

\subsection{Freeze-Out and Decay (FO\&D) }
\label{subsection:FOD}

\begin{figure}[t]\label{fod}
  \center{\includegraphics[scale=0.85]{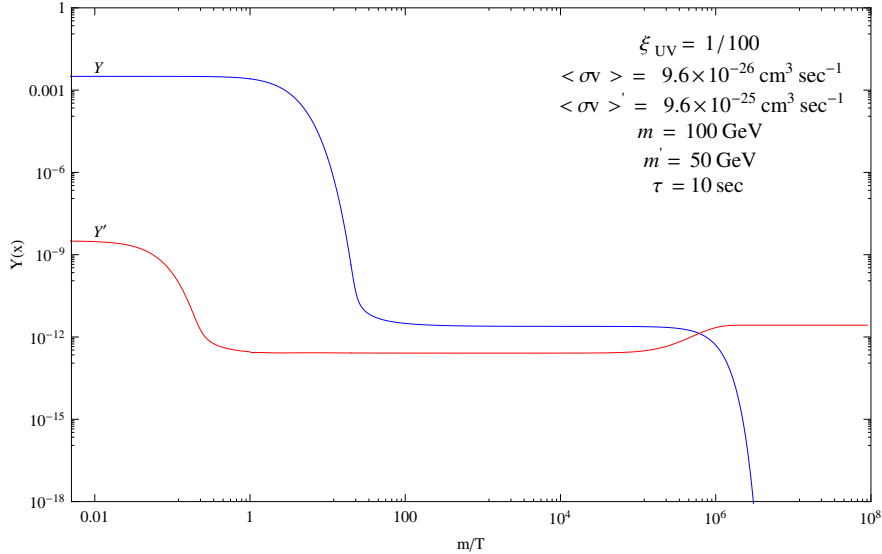}}
\caption{\footnotesize{$X$ and $X'$ yields as a function of $m/T$ for DM production dominated by FO\&D.  The $X$ particles undergo FO and then later decay, yielding an $X'$ abundance that is larger than that which arises from FO$'$.  The parameters $\xi, \sv, m$ and $m'$ are the same as in Figure \ref{fig:FO}, but $\sv'$ is increased giving $Y'_{\FO'} > Y_\FO$.  For $\tau= 1$ second, $X$ decays are occurring at the MeV era.  }}
\label{fig:FOandD}
\end{figure}

In the presence of the portal operator, $\mathcal{O}$,  $X$ is no longer stable.  Thus, after $X$ undergoes FO, it eventually decays into $X'$ particles; we call this DM production mechanism ``Freeze-Out and Decay" (FO\&D).  The resulting $X'$ may form the dominant contribution to the final yield of $X'$, as illustrated in Figure \ref{fig:FOandD}. Assuming the $X$ decay process, $X\rightarrow X' + \ldots $, produces exactly one $X'$ for each $X$, we find   
\be
Y'_\textrm{FO\&D} &=& Y_\FO.
\label{eq:fodeq}
\ee
Consequently, the energy density produced by FO\&D is suppressed relative to that of conventional FO by a factor of $m'/m$.  If FO\&D accounts for the total DM abundance in the universe, then this implies that $\sv = (m'/m)\sv_0$, where recall that $\sv_0$ is the annihilation cross-section needed to account for the measured DM abundance in standard single sector FO. This dilution factor is useful in theories in which FO normally produces an overabundance of DM, for instance as occurs in supersymmetric theories if the LSP is a bino. 

Mechanisms similar to FO\&D have been discussed extensively in the literature for a small subset of candidates for $X$ and $X'$ and operators $\mathcal O$.  In particular, there is a large body of work \cite{Feng:2003uy,Feng:2004zu} concerning the so-called superWIMP scenario in which $X$ is effectively a bino or right-handed slepton and $X'$ is the gravitino.  Axinos \cite{Asaka:2000ew} and goldstini \cite{Cheung:2010mc,Cheung:2010qf} have also been studied as alternative choices for $X'$.

\subsection{Freeze-In (FI) }
\label{subsection:FI}

\begin{figure}[t]\label{lspfi}
  \center{\includegraphics[scale=0.85]{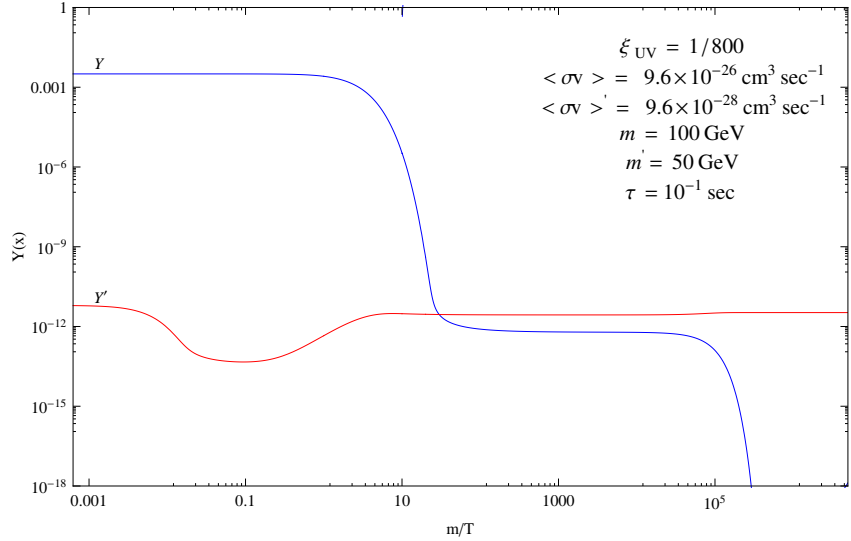}}
\caption{\footnotesize{$X$ and $X'$ yields as a function of $m/T$ for DM production dominated by FI.  
While $X$ is relativistic and in thermal equilibrium, the small fraction of $X$ that decay yield an important contribution to the $X'$ abundance. As soon as $X'$ undergoes FO$'$, the FI mechanism begins to effectively populate an $X'$ abundance that grows until $T$ drops below $m$, when the $X$ abundance becomes exponentially suppressed.  Once the age of the universe reaches $\tau$, the relic $X$ particles from FO all decay but, for the parameter choice for this figure, the increase in $Y'$ from this FO\&D process is sub-dominant to the FI contribution.    }}
\label{fig:FI}
\end{figure}

As the $X$ decay rate is increased, at a certain point a new DM production mechanism, ``Freeze-In'' (FI), begins to dominate.  Here $X'$ particles arise from decays of $X$ particles which are still in thermal equilibrium.   As long as the $X'$ have already undergone FO$'$, the $X'$ produced by FI can comprise the dominant source of DM, as shown in Figure \ref{fig:FI}.  

At any temperature $T>m$ the production of $X'$ by FI generates a yield which goes schematically as
\be
Y'_\FI (T) \propto \Gamma t \propto \frac{\Gamma M_{\rm Pl}}{T^2},
\label{eq:FIT}
\ee
where $t$ is the total time that $X$ is relativistic.
A key aspect of FI by decays is that it is IR dominated by low temperature dynamics; this is true independent of the dimensionality of the connector operator which mediates the decay.  FI can also occur by two-to-two scattering via a marginal coupling (this is also IR dominated). However it turns out to be numerically subdominant compared to that from decays and inverse decays \cite{Hall:2009bx}, so this will not be discussed from now on for simplicity. As $\Gamma$ becomes larger, FI plays an important role in increasing $\xi(T)$ as the temperature drops, as we will discuss in Section \ref{subsection:equil}.  Here we focus on the $X'$ produced after FO$'$.  The FI yield from $X$ decay is dominated by contributions from $T \sim m$ and is the same as computed in \cite{Hall:2009bx} for FI from inverse decays.  The precise formula for the FI yield is
\be
Y'_\textrm{FI} &=&  C_{\rm FI}(x_{\FO'}) \,\, \frac{\Gamma \MPl}{m^2} \\
C_{\rm FI}(x_{\FO'})& \simeq & \frac{135}{2\pi^5}\sqrt{\frac{5}{2}}\frac{g_X}{g_{*S} \sqrt{g_*}} \int^\infty_{x_{\FO'}} K_1(x)x^3 dx \;\;\;
\overset{x_{\FO'} \rightarrow 0}{\rightarrow} \;\;\; C_{\rm FI} =1.64 \frac{g_X}{g_{*S}\sqrt{g_*}},
\label{eq:FI}
\ee 
where $g_X$ is the number of degrees of freedom of $X$ and $x_{\FO'} \equiv m/T_{\FO'}$.  As shown in Eq.~\ref{eq:TFOprime}, in the limit in which the hidden sector is much cooler than the visible sector, $X'$ freezes out very early so  $x_{\FO'}\rightarrow 0$.  Finally, note that if FI accounts for the total DM abundance today, then for weak scale masses this implies a range of lifetimes given by $\tau \simeq 10^{-4}\textrm{ s} - 10^{-1}\textrm{ s}$.  If the decay of $X$ is mediated by a marginal operator with the dimensionless coefficient $\lambda$, then this range of lifetimes corresponds to $\lambda = 10^{-12} - 10^{-11}$.  For decays mediated by a higher dimensional portal interaction, this range applies to $\lambda \equiv (m/M_*)^{d-4}$ where $d$ and $M_*$ are the dimension and scale of the higher dimension operator.  Interestingly, for $d=5$ operators, this corresponds to $M_* \simeq 10^{13}-10^{15}$ GeV, which is roughly of order the GUT scale.

\subsection{Re-Annihilation}
\label{sec:reann}

\begin{figure}\label{lspfi-reann}
  \center{\includegraphics[scale=0.85]{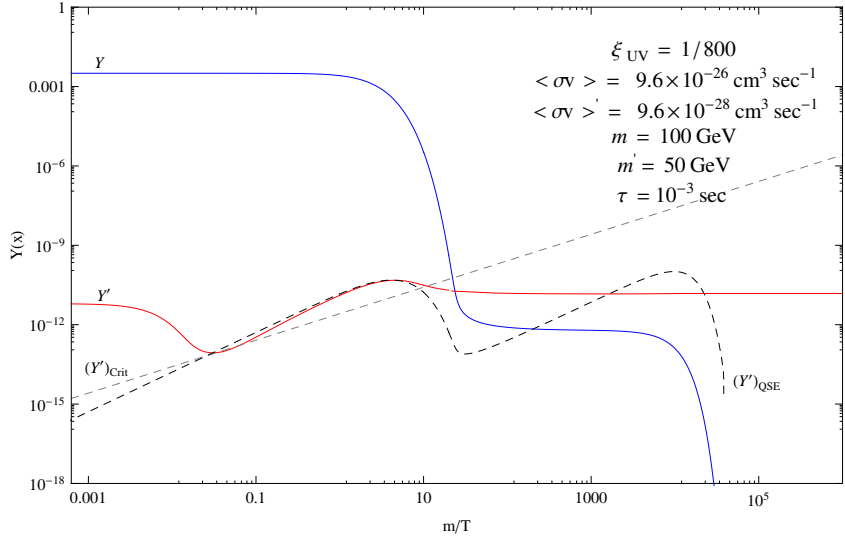}}
\caption{\footnotesize{$X$ and $X'$ yields as a function of $m/T$ for DM production dominated by FI followed by re-annihilation.  Re-annihilation occurs because the FI yield exceeds $Y'_{\rm crit}$ at some temperature where $Y'_{\rm QSE} > Y'_{\rm crit}$.  Once this happens, $Y'$ tracks $Y'_{\rm QSE}$ until it dips below $Y'_{\rm crit}$, which occurs as $Y$ drops rapidly at FO.  The value of $Y'$ at this crossing point is the final yield of $X'$ particles.  }}
\label{fig:FI_ReAnn}
\end{figure}

For simplicity, we have ignored the effects of $X'$ annihilation on FO\&D and FI.  Naively, this is a justifiable omission, since both FO\&D and FI occur only after the hidden sector has undergone FO$'$.  However, even after FO$'$, the $X'$ abundance arising from non-equilibrium production may be so large that the $X'$ annihilation rate grows to exceeds the expansion rate, initiating a new era of $X'$ annihilation that we dub re-annihilation.  Re-annihilation and the resulting $X'$ abundance can be understood through a study of the Boltzmann equation, written in terms of yield variables and $x= m/T$,
\be
x\frac{d}{dx} Y' \simeq -\frac{Y^{\prime 2} }{Y'_{\rm crit}} + \frac{\Gamma Y}{H},
 \label{eq:reann}
\ee
where
\be
Y'_{\rm crit} &\equiv& \frac{H}{\sv' s}.
 \label{eq:Y'crit}
\ee
The first term on the right-hand side of Eq.~\ref{eq:reann} corresponds to $X'$ annihilation; since we are interested in times well after FO$'$,
$Y' \gg Y'_{\rm eq}$ and we ignore inverse annihilations.   The second term is effectively a source  term for $X'$ production, corresponding to the decays of $X$ to $X'$.  At $T \simeq m$ this is the source term which drives FI, while for $T \approx \sqrt{\Gamma M_{Pl}}$ this is the source term which drives FO\&D.  However, the analyses of FI and FO\&D in the previous sections ignored the annihilation term.   

The destruction and production of $X'$ occur faster than the Hubble rate if the first and second terms on the right-hand side of Eq. (\ref{eq:reann}) are larger than $Y'$, respectively, that is if 
\be
Y' > Y'_{\rm crit} \hspace{0.5in} \mbox{\&} \hspace{0.5in} \frac{\Gamma Y}{H} > Y'.
\label{eq:qse-cond}
\ee
In this case the $Y'$ abundance rapidly evolves to a Quasi-Static Equilibrium (QSE) in which the production of $X'$ particles is counter-balanced against depletion from the annihilation process.  This causes the two terms on the right-hand side of Eq.~(\ref{eq:reann}) to cancel, so that $Y'$ becomes equal to $Y'_{\rm QSE}$, where
\be
Y'^2_{\rm QSE} \, = \, \frac{\Gamma Y}{H} Y'_{\rm crit} \, = \, \frac{\Gamma Y}{\sv' s}.
\label{eq:qse-abund}
\ee
Setting $Y' = Y'_{\rm QSE}$ in the first equation of (\ref{eq:qse-cond}), one discovers that QSE is possible only during eras having
\be
Y'_{\rm QSE} > Y'_{\rm crit}.
\label{eq:qse-cond2}
\ee

\begin{figure}\label{fod-reann}
  \center{\includegraphics[scale=0.85]{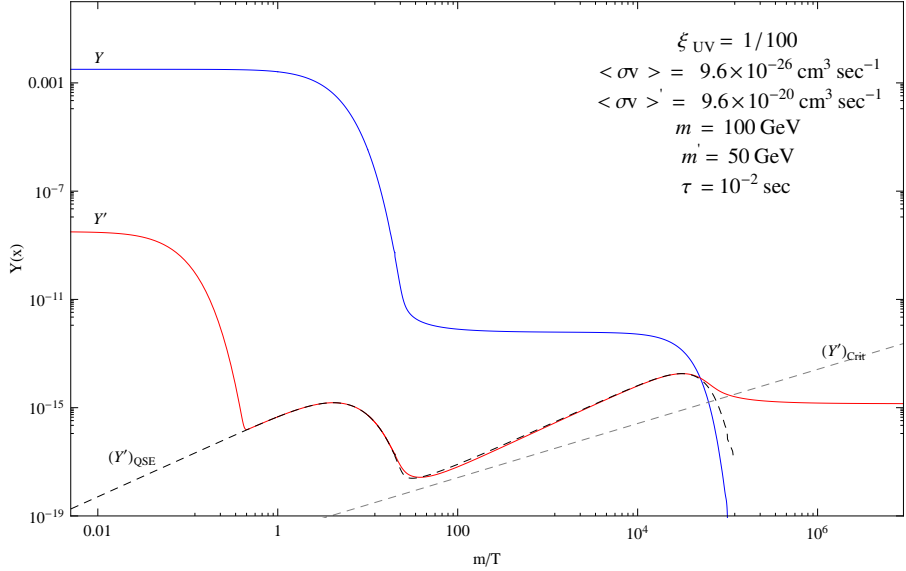}}
\caption{\footnotesize{$X$ and $X'$ yields as a function of $m/T$ for DM production dominated by FO\&D followed by re-annihilation.  $Y'$ tracks $Y'_{\rm QSE}$ until it drops below $Y'_{\rm crit}$, which occurs when $Y$ drops sharply as $X$ decay.  }}
\label{fig:FOandD_ReAnn}
\end{figure}

As $X'$ undergo FO$'$, the depletion of $X'$ will stop once $Y'$ drops to $Y'_{\rm QSE}$, provided Eq. (\ref{eq:qse-cond2}) is satisfied, as shown in Figures \ref{fig:FI_ReAnn} and \ref{fig:FOandD_ReAnn} which were produced by numerically solving the exact Boltzmann equations.  Subsequently $Y'$ tracks $Y'_{\rm QSE}$ until Eq. (\ref{eq:qse-cond2}) is violated.  This always eventually happens because $Y'_{\rm crit}$ grows linearly with $m/T$ and $Y'_{\rm QSE}$ drops as $Y$ is reduced by FO or $X$ decay.
When QSE ends, i.e.~when $Y'_{\rm QSE} = Y'_{\rm crit}$, the re-annihilation rate drops below the expansion rate so that QSE is lost and the $X'$ yield becomes constant.  The final $X'$ abundance is fixed by the value of $Y'$ at this point.  We will denote FO\&D and FI which are subsequently followed by re-annihilation by FO\&D$_{\rm r}$ and FI$_{\rm r}$.   Note that re-annihilation did not occur in the examples shown in Figures \ref{fig:FO}-\ref{fig:FI} because  Eq. (\ref{eq:qse-cond2}) was violated at FO$'$ and all subsequent eras.

Figures~\ref{fig:FI_ReAnn} and \ref{fig:FOandD_ReAnn} show yield plots for cases where DM is dominated by FI$_{\rm r}$ and  FO\&D$_{\rm r}$, respectively.  The dashed lines indicate $Y'_{\rm QSE}$ and $Y'_{\rm crit}$.  In both plots, one sees that, once inverse annihilations of $X'$ can be neglected, QSE occurs during eras with $Y'_{\rm QSE} > Y'_{\rm crit}$, with $Y'$ accurately tracking $Y'_{\rm QSE}$.  Eventually, $Y'_{\rm QSE}$ falls below $Y'_{\rm crit}$ and QSE ends.  The final DM abundance reflects the value of $Y'_{\rm QSE}$ at the end of the QSE era, which we now study.

To analytically compute $Y'_{\rm FO\&D_{\rm r}}$ and $Y'_{\FI_{\rm r}}$, we must compute the temperature at the end of QSE, $T_{\rm r}$, which is by definition the solution to the equation
\be
Y'_{\rm QSE}(T_{\rm r}) &=& Y'_{\rm crit} (T_{\rm r}).
\ee  
According to Eq.~(\ref{eq:qse-abund}), $Y'_{\rm QSE} \propto \sqrt{Y}$, so the solution to this equation depends on the expression for $Y$ during the era under consideration.  Going from $T$ to $x$ variables, then for FO\&D$_{\rm r}$ and FI$_{\rm r}$ we must solve the transcendental equations
\be 
\label{xfirfodr}
\left\{
\begin{array}{cc}
x_{\rm FO\&D_{\rm r}}^{1/2} \, e^{-x_{\rm FO\&D_{\rm r}}}=\frac{1}{2} \left(\frac{5}{18 \pi^2 g_{\star}} \right)^{1/4} \,\frac{1}{Y_{\rm FO}\MPl^{3/2}  \Gamma^{1/2} \sv' } ,& \quad {\rm FO\&D}_{\rm r}  \\
x_{{\rm FI}_{\rm r}} ^{5/2} \, e^{-x_{{\rm FI}_{\rm r}} }= \frac{\sqrt{2} \pi^{7/2}}{45} \frac{g_*}{g}\frac{m}{\MPl^2 \Gamma \sv' },  & \quad {\rm FI}_{\rm r} 
\end{array}
\right.
\ee
Hence, the final yields for FO\&D and FI followed by re-annihilation are given by
\be
Y'_{\rm FO\&D_{\rm r}} &=& Y'_{\rm crit} (T_{\rm FO\&D_{\rm r}}) \\
Y'_{\rm FI_{\rm r}} &=& Y'_{\rm crit} (T_{\rm FI_{\rm r}}),
\ee  
This result is very similar in structure to the yield obtained from standard FO$'$, namely $Y'_{\rm FO'} = Y'_{\rm crit} (T_{\rm FO'})$.  For $s$ wave annihilation $Y'_{\rm crit} \propto 1/ T$, so the re-annihilation yields obey the simple relation
\be
T_{\rm FO\&D_{\rm r}} Y'_{\rm FO\&D_{\rm r}} = T_{\rm FI_{\rm r}} Y'_{\rm FI_{\rm r}} = T_{\rm FO'} Y'_{\rm FO'}
\ee  
To be concrete, this implies that the DM yield for ${\rm FO\&D}_r$ and $\FI_r$ are given by the formulas
\be \label{fodr-fir}
Y'_{{\rm FO\&D_r}} &\simeq& \frac{3}{2 \pi} \sqrt{\frac{5}{2}} \frac{\sqrt{g_*}}{g_{* S}} \frac{1}{\MPl\,\sv^{\prime}}\frac{1}{T_{\rm FO\&D_r}} \nonumber\\
Y'_{{\FI_r}} &\simeq& \frac{3}{2 \pi} \sqrt{\frac{5}{2}} \frac{\sqrt{g_*}}{g_{*S}} \frac{1}{\MPl\,\sv^{\prime}}\frac{1}{T_{\FI_r}}, 
\ee where $T_{\FI_r}=m/x_{\FI_r}$, and $T_{\rm FO\&D_r}=T_{\text{Decay}}/\sqrt{x_{\rm FO\&D_r}}$ so that the exponential in Eq.~(\ref{xfirfodr}) goes simply as $e^{-x}$. Here $T_{\rm Decay}$ is the temperature at which $X$ decays and $x_{\FI_r}$ and $x_{\rm FO\&D_r}$ are given by the solutions of Eq.~(\ref{xfirfodr}).
 
As we have seen, the DM yield from FO\&D and FI can differ substantially from FO\&D$_{\rm r}$ and FI$_{\rm r}$.  The condition for avoiding re-annihilation effects  is $Y'_{\rm FI} < Y'_{\rm crit}(T = m)$ for FI and $Y_{\rm FO} < Y'_{\rm crit}(T = \sqrt{\Gamma M_{\rm Pl}})$ for FO\&D.

\subsection{Sector Equilibration}
\label{subsection:equil}
  
Thus far we have ignored the effects of the connector operator $\mathcal O$ on the thermal properties of the visible and hidden sectors.  Specifically, there is the danger that $\mathcal O$ couples the sectors so strongly that they actually come into thermal equilibrium.  This scenario corresponds to the case where $\xi = T'/T \approx 1$ near the weak scale.  In general, $\xi$ is temperature dependent and receives contributions from UV and IR-sensitive physics,
\be\label{eq:xi2}
  \xi^4(T) = \xi_{\rm UV}^4 + \xi_{\rm IR}^4(T).
\ee
The UV contribution to the hidden sector temperature, $\xi_{\rm UV}$, arises from two sources, so $\xi_{\rm UV}^4 = \xi^4_{\rm inf} + \xi^4_R$.  If the inflaton couples directly to the hidden sector, then $\xi_{\rm inf}$ is generated by an initial heating of the hidden sector from inflaton decays.  This contribution is independent of the portal interactions, and was discussed in earlier sections.  On the other hand, $\xi_R$ results from scattering processes mediated by $\mathcal{O}$ occurring at the reheat temperature, $T_R$.  These processes are active if $\mathcal O$ is a higher-dimension operator.  In this case, $\cal O$ contributes a 2-to-2 scattering cross-section, $\sv_R$, which produces an $X'$ yield of 
\be 
Y'_R \sim \MPl T_R \sv_R.
\label{eq:reheatyield}
\ee
The $T_R$ dependence of $\sv_R$ depends on the dimensionality of  $\mathcal O$.
If $X'$ is inert, $Y'_R$ can easily overclose the universe unless $T_R$ is sufficiently small.  For instance, in the case of gravitino LSP this is the origin of the well-known bounds on $T_R$ from overclosure \cite{Moroi:1993mb}.  On the other hand, our assumption is that $X'$ possesses self-interactions, so $X'$ particles produced by scattering at reheating will be efficiently thermalized via the $X'$ annihilation until the onset of FO$'$.  Only after FO$'$ can an abundance $X'$ particles be produced via FI.  Consequently, in the presence of $X'$ annihilations, the FI abundance from the higher dimension operator $\mathcal{O}$ is given by Eq.~\ref{eq:reheatyield}, only with $T_R$ replaced by $T_{\rm FO'}$.  Because $T_{\rm FO'}$ is not exceedingly far from the weak scale, this UV dominated FI contribution from 2-to-2 scattering will in general be subdominant to the IR dominated FI contribution from decays discussed in Section \ref{subsection:FI}.

Since the $X$ particles are produced with energy $T_R$, the hidden sector is reheated by the visible sector to an energy density given by $T_R^{\prime 4} \sim Y'_R T_R^4$.  Thus, the ratio of visible and hidden sector temperatures is given by
\be
\xi_R &\sim& (\MPl T_R \sv_R)^{1/4},
\label{eq:xiR}
\ee
in the case where there is 2-to-2 scattering processed mediated by higher dimension operators.

Finally, let us consider the IR contribution to $\xi$, which essentially arises from FI.  For decays of $X$ at temperature $T$, FI produces a yield $Y'_\FI$ given by (\ref{eq:FIT}).  The produced $X'$ particles have an energy distribution characteristic of temperature $T$.  We assume that the interactions of the hidden sector are sufficient to rapidly thermalize the energy of these $X'$ into distributions of all the hidden sector particles at temperature $T'$.  As long as the hidden sector remains sufficiently cool, then FI will be mediated by decays and inverse decays can be ignored.  From this point of view, FI leaks energy and entropy out of the visible sector into the hidden sector.  This effect is especially important when the lifetime of $X$ is short, since the strength of this energy leakage is proportional to $\Gamma$.  The contribution to $\xi_{\rm IR}$ from FI is  
\be 
\xi_{\rm IR}^4(T)  &=& A \;  \frac{M_{\rm Pl} \Gamma}{T^2} \hspace{.5in}  (T>m), 
\label{eq:Aeq}
\ee
with an analytic estimate giving $A \approx 135 \sqrt{5} \,\,  g_X /(\pi^5 \, \sqrt{g_*} g'_*)$.   As $T$ drops below $m$ the FI process gets exponentially switched off, so
\be
\xi_{\rm IR}(T) \simeq \xi_{\rm IR}(m)    \hspace{.5in}  (T<m).
\ee
As $\Gamma$ increases we reach a critical point where $\xi_{\rm IR}(T\simeq m)=1$; the FI process is now so strong that the two sectors are equilibrated when $T\simeq m$.  Since $m>m'$ the $X$ particles no longer undergo FO, and instead track their equilibrium abundance by decaying.  Consequently, there is single thermalized sector and DM results from the single process of FO$'$.  For $\xi_{\rm UV}^4 \ll 1$, the critical lifetime that leads to this equilibration scales as $m^{-2}$
\be
\tau_{\rm min} \simeq 10^{-13} \mbox{ s} \;  \left( \frac{100 \, \mbox{GeV}}{m}\right)^2 \left( \frac{100}{g'_*(T \simeq m)/g_X} \right),
\label{eq:taumin}
\ee
where the numerical coefficient has been extracted from our numerical results.
Since the contribution to the heating of the hidden sector from FI is model-independent, we include its effects throughout our analysis.  On the other hand, the UV-sensitive contributions are very model-dependent, so we take $\xi_{\rm UV}$ to be a free parameter which is small.

\subsection{Summary of Results}\label{subsection:summarize}

Let us now summarize the results of the entire section. We have shown that in our setup the present day abundance of DM particles can originate only from a handful of cosmological production mechanisms.  In the simplest case, the hidden sector  undergoes FO$'$, yielding a thermal relic abundance of $X'$ particles.  Alternatively, via the FO\&D mechanism, an abundance of $X$ particles can FO in the visible sector, and then decay very late into $X'$ particles.  The FI mechanism functions so that $X$ particles, while still in thermal equilibrium, provide an abundance of $X'$ particles through decays.  Lastly, if the $X'$ yields from FO\&D and FI exceed a certain critical yield, $Y'_{\rm crit}$, then the hidden sector enters an era of re-annihilation.  The final abundances for FO\&D$_{\rm r}$ and FI$_{\rm r}$ are controlled by the temperature at the end of this re-annihilation era.
The analytic expressions for the DM yield are 
\be
Y_{\rm FO\&D}^\prime &=& C_{\rm FO} \; \frac{1}{\MPl \sv} \frac{1}{T_{\rm FO}} \; \propto \; \frac{1}{m \langle \sigma v\rangle }  \\
Y_{\rm FO^{\prime}}^{\prime} &=& C_{\rm FO} \; \frac{1}{\MPl \sv^{\prime}} \frac{1}{T_{\rm FO^{\prime}}} \; \propto \; \frac{\xi_{\rm FO'}}{m' \langle \sigma v \rangle'}  \\
Y_{\rm FO\&D_{r}}^{\prime} &=& C_{\rm FO} \; \frac{1}{\MPl \sv^{\prime}} \frac{1}{T_{\rm FO\&D_{r}}}  \; \propto \; \frac{\sqrt{\tau }}{\langle \sigma v \rangle'}\\
Y_{\rm FI_{r}}^{\prime} &=& C_{\rm FO} \; \frac{1}{\MPl \sv^{\prime}} \frac{1}{T_{\rm FI_{r}}} \; \propto \; \frac{1}{m \langle \sigma v\rangle '}  \\
Y_{\rm FI}^{\prime} &=& C_{\rm FI} \; \frac{\Gamma \MPl}{m^2} \; \propto \; \frac{1}{\tau m^2} 
\ee
in the cases where the various mechanisms dominate the contribution to the total yield. 
Here we have defined the dimensionless constants $C_{\rm FO}=\frac{3}{2 \pi} \sqrt{\frac{5}{2}} \frac{\sqrt{g_*}}{g_{*S}}$ and 
$C_{\rm FI} = 1.64 \frac{g}{g_{*S} \sqrt{g_*}}$, and the various temperatures are defined in Table \ref{yield-exp}.  The dimensionless values $``x"$ defined in the second column of this table are determined in each case by solving a transcendental equation of the general form: \be x^n\,e^{-x} = f(m,m',\sv,\sv^{\prime},\xi, \tau)\ee for some rational number $n$ and where $f$ is a some function of the arguments. Here we have taken an approximate solution in which the effect of $x^n$ is neglected.

\begin{table}[h!]
\centering
\small\addtolength{\tabcolsep}{-5pt}
\begin{center}
\begin{tabular}{||c|c||}
\hline \hline
   T & $x$ \\
\hline
\hline
$\TFO = \frac{m}{x_{\rm FO}}$ 
& $x_{\rm FO} \sim \ln \left[\frac{\sqrt{45}}{2} \frac{1}{\pi^{5/2}} \frac{g}{\sqrt{g_{\star}}} 
\MPl m \sv \right]$ \\ \cline{1-2}
$T_{\rm FO^{\prime}} = \frac{1}{\xi_{\FOp}} \frac{m^{\prime}}{x'_{\rm FO^{\prime}}}$ 
& $x'_{\rm FO^{\prime}} \sim \ln \left[ \frac{\sqrt{45}}{2} \frac{1}{\pi^{5/2}} 
\frac{g}{\sqrt{g_{\star}}}\xi_{\FOp}^2
\MPl m^{\prime} \sv^{\prime} \right]$ \\ \cline{1-2}
$T_{\rm FO\&D_{r}} = \frac{T_{\text{Decay}}}{\sqrt{x_{\rm FO\&D_{r}}}} = 
\left(\frac{45}{2 \pi^2 g_{\star}} \right)^{\frac{1}{4}} \sqrt{\frac{\MPl \Gamma}{x_{\rm FO\&D_{r}}}}$ 
& $x_{\rm FO\&D_{r}} \sim \ln \left[ \left( \frac{90}{\pi^2 g_{\star}} \right)^{\frac{1}{4}} \sqrt{2} 
x_{FO} \frac{\sv^{\prime}}{\sv} \frac{\sqrt{\Gamma \MPl}}{m}
\right]$  \\ \cline{1-2}
 $T_{\rm FI_{r}} = \frac{m}{x_{FI_{r}}}$ 
& $x_{\rm FI_{r}} \sim \ln \left[ \frac{45}{\sqrt{2} \pi^{7/2}} \frac{g}{g_{\star}} 
\frac{\MPl^2 \sv^{\prime} \Gamma}{m}\right]$ \\ \cline{1-2}
\hline\hline
\end{tabular}
\end{center}
\caption{\footnotesize{Expressions for the various temperatures relevant for each DM production mechanism.  The $``x''$ quantities employed in the first column are given approximate expressions in the second column. The quantity $\xi_{\FOp}$ is computed in (\ref{xifop}) below. }}.
\label{yield-exp}
\end{table}

Note that only $\FOp$ depends on $\xi$; in particular it depends on the value of $\xi$ at $T_{\FOp}$ which is denoted in the Table \ref{yield-exp} as $\xi_{\FOp}$. From the analysis in Section \ref{subsection:equil}, one notes that the quantity $\xi_{\FOp} \equiv \xi(T_{\FOp})$ has  different forms depending on whether 
$T_{\FOp}$ is greater or smaller than $m$. $\xi_{\FOp}$ can be computed as: 
\be \label{xifop}
\xi_{\FOp} &=& 
\left\{
\begin{array}{ll}
\left(\xi_{\rm UV}^4+\frac{A\Gamma\MPl}{m^2}\right)^{1/4} &,\quad  T_{\FOp}< m\\
\left(\frac{A\,\Gamma\MPl\,x_{\rm FO}^2}{2\,m'^2}\right)^{1/2}\,\left[1+\left(1+\frac{4\,\xi_{\rm UV}^4\, m'^4}{A^2\Gamma^2\,\MPl^2\,x_{\rm FO}^4}\right)^{1/2}\right]^{1/2} &,\quad  T_{\FOp} > m 
\end{array}
\right.
\ee
where $A$ is as defined after Eq.~(\ref{eq:Aeq}).

\section{Cosmological Phase Diagrams}\label{cpd}

A primary aim of this paper is to  identify  and characterize all possible mechanisms of DM production which can arise within our general two-sector framework.  To this end, we have simulated the cosmological history of this system over a broad range of values for the relevant parameters:
\be
\{m, m',  \langle\sigma v\rangle,\langle \sigma v \rangle ', \xi_{\rm UV},\tau\},
\ee
where  $\xi_{\rm UV}$ is the UV initial condition for $\xi$ which receives contributions from the decay of the inflaton as well as scattering processes from higher-dimensional operators described in Eq.~(\ref{eq:xi2})\footnote{Here we also take $\xi_{\rm UV}$ to include effects from additional sources of entropy dumping into either sector before the weak era, so that $\xi_{\rm UV}$ is effectively the weak scale value of $\xi$, modulo the contribution from $X$ decays in the IR.}.  As noted earlier, it is quite remarkable that the cosmology is determined solely by just a handful of quantities.

\begin{figure}[h!]
  \center{\includegraphics[scale=1.0]{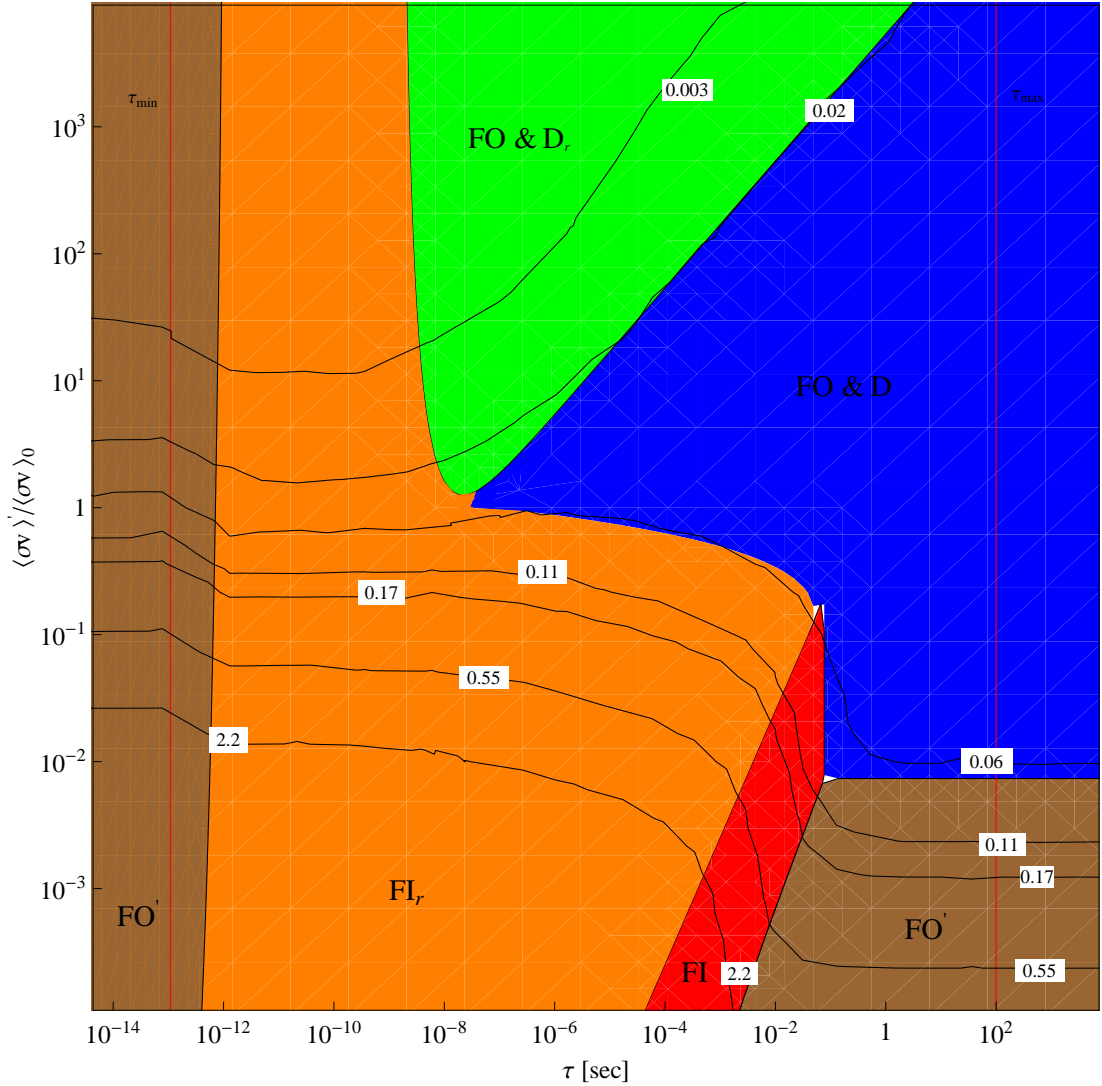}}
\caption{\footnotesize{Cosmological phase diagram showing regions in the $\frac{\langle \sigma v \rangle'}{\langle \sigma v \rangle_{0}}$ versus $\tau$ plane where different mechanisms contributing to the relic abundance dominate. The values of the other relevant parameters are chosen as: $\xi_{\rm UV}=0.01,\,m=100\,\mathrm{GeV},\,m'=50\,\mathrm{GeV},\,\sv =\sv_0=3\times 10^{-26} \textrm{ cm}^3/\textrm{s}$. Contours of $\Omega h^2$, computed from a full numerical analysis for $Y'$ (numerical solution of Eq.~(\ref{eq:Bn}) and Eq.~(\ref{eq:Bn'})), are shown. Regions in which the various mechanisms dominate are shown in different colors. These regions are computed analytically and overlaid on the numerical plot. The agreement is quite good.}}
\label{phasediag}
\end{figure}

In this section we present a series of ``cosmological phase diagrams"  depicting the regions in parameter space in which each mode of DM production, i.e.~FO\&D, FI, etc., accounts for the dominant contribution to the present day DM abundance.  For example, Figure \ref{phasediag} is a cosmological phase diagram in the $\tau - \frac{\langle \sigma v\rangle ' }{\langle \sigma v \rangle}$ plane for  particular values of $\{m, m', \sv, \xi_{\rm UV}\}$, as explained in the caption.  Each colored shaded region corresponds to a particular DM production mode which is dominant in that region.  The boundaries of each phase have been computed analytically.  The solid black contours correspond to the present day DM relic abundance, calculated numerically using the coupled Boltzmann equations in Eqs.~(\ref{eq:Bn}).   Let us consider some of the features of Figure \ref{phasediag} in detail.

First, note that the lifetime $\tau$ is constrained both from above and below as shown by the red vertical lines in Figure \ref{phasediag}. The upper limit is easy to understand as it originates from the requirement the decay products of $X$ do not ruin the successful predictions of BBN \cite{Moroi:1993mb}. While the precise constraint depends on the nature of $X$ and its decay products, here we use
\be \tau_{\rm max}\simeq 100\textrm{ s}, \ee   
as depicted by the red line at large $\tau$ in Figure \ref{phasediag}. The lower limit on $\tau$ comes from demanding that the two sectors are \emph{not} in thermal equilibrium with each other at the weak scale. In Section \ref{subsection:equil} we found that for $\xi_{\rm UV}^4 \ll 1$, the two sectors becomes thermally equilibrated if the lifetime is shorter than
\be \tau_{\rm min}  \simeq 10^{-13}\,{\rm s}\,\left(\frac{100\,{\rm GeV}}{m}\right)^2  \left( \frac{100}{g'_*(T \simeq m)/g_X} \right). \ee
For $\tau \lesssim 10^{-13}$ s, the two sectors are, from a cosmological perspective, a single sector.  As a consequence, there is no distinction between FO and FO$'$---the visible and hidden sectors together maintain thermal equilibrium until $X'$ undergoes single sector FO$'$. This is depicted by the {\it brown} region in the left in Figure \ref{phasediag}. The yield $Y'_{\FOp}$ in this region is independent of $\tau$ because $\xi_{\FOp}=1$ in the expression for $Y'_{\FOp}$ in Table \ref{yield-exp}. Note that the {\it brown} region extends to $\tau$ larger than $\tau_{\rm min}$, as will be discussed soon.  

Next, let us discuss the salient features of Figure \ref{phasediag} as we move from the largest to the smallest allowed values for $\tau$. Furthermore, for a given value of $\tau$, the dominant DM production mechanism changes as a function of $\frac{\sv^{\prime}}{\sv}$. For example, for the choice of parameters in Figure \ref{phasediag}, one finds that FO\&D and $\FOp$ dominate over FI for $10^{-1}\textrm{ s}\lesssim \tau \lesssim 100$ s. This is because the lifetime is too long (the coupling is too weak) for the FI mechanism to set in, as shown in Section \ref{lspfi}. For values of $\tau$ in the above range with $\frac{\sv^{\prime}}{\sv} \gtrsim 10^{-1}$, corresponding to the {\it blue} shaded region, FO\&D is the dominant mechanism. Indeed, from the shape of the contours one can see that in this region the relic abundance is independent of $\frac{\sv^{\prime}}{\sv}$ and $\tau$ as dictated by the expression for $Y'_{{\rm FO\&D}}$ in Table \ref{yield-exp}. For $\frac{\sv^{\prime}}{\sv} \lesssim 10^{-2}$, $\FOp$ starts to dominate over FO\&D since the annihilation cross-section of $X'$ becomes sufficiently small. This is shown by the {\it brown} region in the right in the figure. For such long lifetimes, one finds that $\xi_{\FOp} \approx \xi_{\rm UV}$ from (\ref{eq:xi2}), implying that $Y'_{\FOp}$ is essentially independent of $\tau$ (see the expression for $Y'_{\FOp}$ in Table \ref{yield-exp}). Thus, the contour lines in this region are roughly horizontal. The boundary between the FO\&D and $\FOp$ regions for this range of $\tau$ is given by :
\be \label{fod-fop}
Y'_{{\rm FO\&D}} = Y'_{\FO^{\prime}}\implies \,
\frac{\sv^{\prime}}{\sv} &=&\xi_{\FOp} \frac{m}{m^{\prime}}\nonumber\\
\frac{\sv^{\prime}}{\sv} &\propto& \text{constant}
\ee which in the $\frac{\sv^{\prime}}{\sv}-\tau$ plane corresponds to $\frac{\sv^{\prime}}{\sv} \approx$ constant, since $\xi_{\FOp} \approx \xi_{\rm UV}$.

For $10^{-8}\textrm{ s}\lesssim \tau \lesssim 10^{-4}-10^{-1}$ s, FI and $\mathrm{FI_{r}}$ begin to dominate over FO\&D and $\FOp$ because the portal interactions between  sectors is growing stronger. In particular, FI is dominant in the narrow band shown by the {\it red} shaded region in Figure \ref{phasediag}.  For FI to be the dominant mechanism, the yield from FI must be larger than that from $\FOp$ but smaller than that from $\FI_{r}$; hence the narrow band in which FI is the dominant mechanism. The boundary curves between the $\FI$ region and the $\FOp$ $\&$ $\FI_r$ regions are given by:  
\be\label{fi-fop-fir}
Y'_{{\FI}} = Y'_{\FI_r}\implies\, \frac{\Gamma \MPl^2 \sv^{\prime}}{m} &=& \frac{C_{\rm FO}}{C_{\rm FI}} x_{\FI_{r}}
\nonumber\\
\frac{\sv^{\prime}}{\sv} &\propto& \tau\\
Y'_{{\FI}} = Y'_{\FOp}\implies \, \frac{\Gamma \MPl^2 \sv^{\prime}m'}{\xi_{\FOp}\,m^2} &=& \frac{C_{\rm FO}}{C_{\rm FI}} x_{\FOp}
\nonumber\\
\frac{\sv^{\prime}}{\sv} &\propto& \tau \label{eq:FI=FO'}
\ee Both curves imply a linear relation between $\frac{\sv^{\prime}}{\sv}$ and $\tau$, only with different coefficients\footnote{The boundary shown in Eq.~(\ref{eq:FI=FO'}) depends on $\xi_{\FOp}$ as well, but in this region $\xi_{\FOp} \approx \xi_{\rm UV}$ is a constant.}. Also, note from Table \ref{yield-exp} that $Y'_{\FI}$ does not depend on $\frac{\sv^{\prime}}{\sv}$, hence the contour lines for the relic abundance are almost vertical. They are not completely vertical since it turns out that $\FI$ never fully dominates the contribution to the relic abundance.  

For $10^{-8} \textrm{ s} \lesssim \tau \lesssim 10^{-4}-10^{-1}$ s  with $\frac{\sv^{\prime}}{\sv} \lesssim 1$, the FI yield becomes so large that it exceeds the critical yield at $T_{\FI_r}$, implying that $X'$ starts to re-annihilate and the yield is given by $Y'_{\FI_r}$. This is shown by the {\it orange} region in the figure. From Table \ref{yield-exp}, one sees that contours for $Y'_{\FI_r}$ are essentially horizontal since the dependence on $\tau$ only arises from the logarithm. As one increases $\frac{\sv^{\prime}}{\sv}$ above roughly unity, FO\&D starts to dominate over $\FI_{r}$ as shown by the {\it blue} region, since the annihilation cross-section of $X'$ becomes large enough that the yield $Y'_{\FI_r}$ becomes smaller than $Y'_{\mathrm{FO\&D}}$. The boundary curve between these two regions is defined by: 
\be \label{fod-fir}
Y'_{{\rm FO\&D}} = Y'_{\FI_{r}}\implies\,
\frac{\sv^{\prime}}{\sv} &=& \frac{x_{\rm FI_{r}}}{x_{\rm FO}}\nonumber\\
\frac{\sv^{\prime}}{\sv} &\propto& x_{\rm FI_{r}} \propto \ln \left[\frac{\sv^{\prime}}{\sv} \frac{1}{\tau}\right]
\ee showing that $\frac{\sv^{\prime}}{\sv}$ is essentially constant up to a logarithmic dependence.

If $\frac{\sv^{\prime}}{\sv}$ is further increased, FO\&D is eventually superseded by $\mathrm{FO\&D_{r}}$ as shown by the {\it green} region. In this case the $X'$ yield from the freezeout and decay of $X$ is so large that it becomes larger than the critical yield; so $X'$ starts re-annihilating with a yield given by $Y'_{\mathrm{FO\&D}_r}$. From Table \ref{yield-exp}, the contour plots for $Y'_{\mathrm{FO\&D}_r}$ follow a simple power law ($\frac{\sv^{\prime}}{\sv}\propto \sqrt{\tau}$). The boundary curve between the FO\&D and ${\rm FO\&D_r}$ regions is given by: 
\be\label{fod-fodr}
Y'_{{\rm FO\&D_{r}}} = Y'_{{\rm FO\&D}}\implies \,
\frac{\sv^{\prime}}{\sv} &=& \frac{1}{A_{\rm FO\&D_{r}}} \frac{\sqrt{x_{\rm FO\&D_{r}}}}{x_{\rm FO}} 
\frac{m}{\sqrt{\Gamma \MPl}} \nonumber\\
\frac{\sv^{\prime}}{\sv} &\propto& \sqrt{\tau}
\ee showing that the boundary also follows the same power law as the DM yield contours.

For ${\rm few}\times 10^{-13} \textrm{ s}\lesssim \tau \lesssim 10^{-8}$ s, $\FI_r$ is again the dominant mechanism as shown by the {\it orange} region. It dominates over 
$\mathrm{FO\&D_r}$ in particular, since the yield is inversely proportional to the relevant temperatures for $\mathrm{FO\&D_r}$ and $\FI_r$, which are $T_{{\rm FO\&D_r}}$ and $T_{\FI_r}$ respectively, and $T_{\FO_r}$ is larger for small lifetimes (large $\Gamma$) compared to $T_{\FI_r}$ from Table \ref{yield-exp}.  The boundary curve between these two regions is given by:
\be\label{fodr-fir}
Y'_{{\rm FO\&D_r}} = Y'_{\FI_r}\implies\, \sqrt{\frac{\Gamma \MPl}{x_{\rm FO\&D_r}}}\,A_{\rm FO\&D_r} &=& \frac{m}{x_{\rm FI_r}}
\nonumber\\
\tau \propto \frac{x_{\rm FI_r}^2}{x_{\rm FO_r}} &\propto& \frac{\left(\ln\left[\frac{\sv^{\prime}}{\sv}\frac{1}{\tau}\right]\right)^2}{
\ln\left[\frac{\sv^{\prime}}{\sv}\frac{1}{\sqrt{\tau}}\right]}
\ee which is roughly $\tau \propto$ constant, although more precisely there is a  dependence on $\frac{\sv^{\prime}}{\sv}$ and $\tau$ from the logarithms.

Finally, for $\tau_{\rm min} < \tau \lesssim {\rm few}\times \tau_{\rm min}$, $\FOp$ again dominates for the entire range of $\frac{\sv^{\prime}}{\sv}$ as is shown by the {\it brown} region in Figure \ref{phasediag}. To understand this, it has to be compared to the next relevant mechanism, namely $\FI_{\rm r}$. Since the lifetime $\tau$ is short in this region, $\xi_{\FOp}$, which is given by the expression in the top line in Eq.~(\ref{xifop}) in this case, becomes large and close to unity. Hence, in this region, $\xi_{\FOp} > \frac{m'}{m}\frac{x_{\rm FI_r}}{x_{\rm FO'}}$, or equivalently $T_{\FOp} < T_{\FI_{\rm r}} < m$ from Table \ref{yield-exp}. Since $\sv$ has been fixed in Figure \ref{phasediag}, from Table \ref{yield-exp} and (\ref{xifop}), the contours for the DM yield in this region have a dependence $\frac{\sv^{\prime}}{\sv} \propto \tau^{-1/4}$. The boundary curve between the two regions is given by:
\be\label{fop-fir}  
Y'_{\FOp} = Y'_{\FI_{\rm r}}\implies\, \frac{1}{\xi_{\FOp}}\frac{m'}{x_{\rm FO'}} &=& \frac{m}{x_{\rm FI_r}}\nonumber\\ 
\tau &\propto& \frac{1}{\left(\ln\left[\frac{\sv^{\prime}}{\sv}\frac{1}{\tau}\right]\right)^4}
\ee which is again approximately $\tau \propto$ constant, but more pecisely with an additional dependence on $\frac{\sv^{\prime}}{\sv}$ and $\tau$ in the logarithm.

\subsection{Behavior of Phase Space Diagram}

\begin{figure}[t]
\centering
\begin{tabular}{ccc}
\includegraphics[scale=0.65]{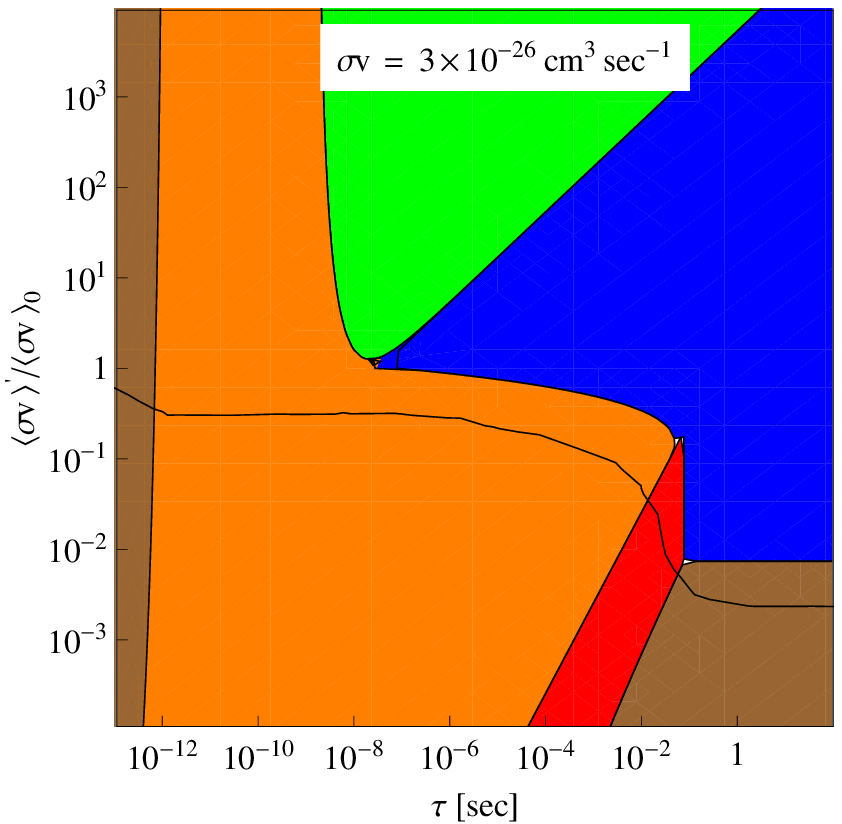}& 
\includegraphics[scale=0.65]{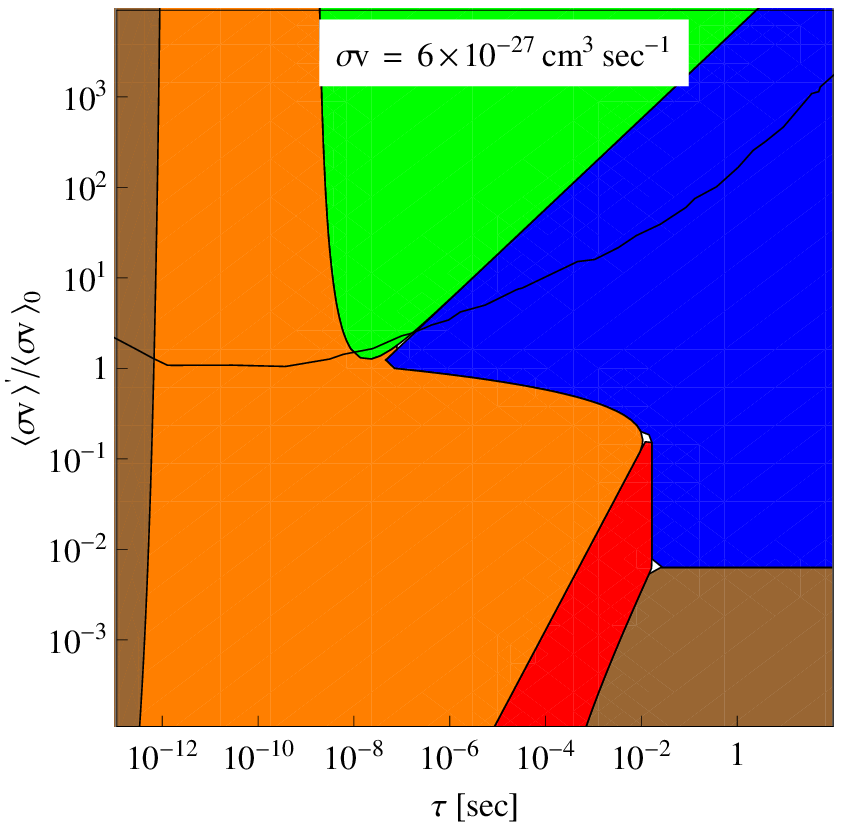}& 
\includegraphics[scale=0.65]{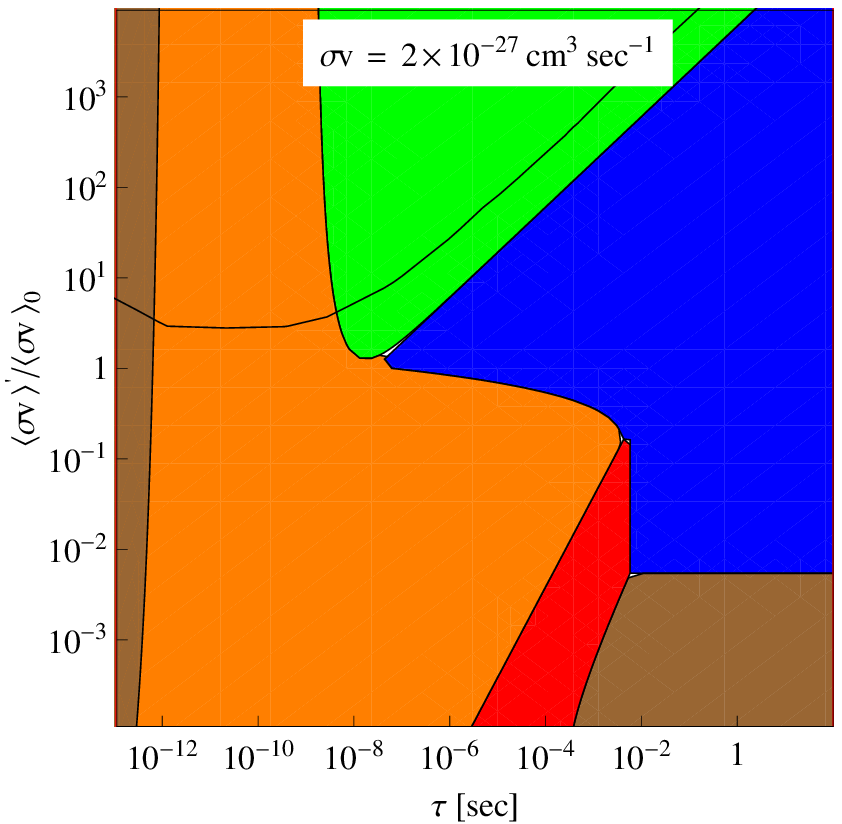}\\
\end{tabular}
\caption{\footnotesize{Sequence of cosmological phase diagrams with $\sv$ varying.  Other parameters are as in Figure \ref{phasediag}:  $(\xi_{\rm UV}=0.01,\,m=100\,\mathrm{GeV},\,m'=50\,\mathrm{GeV})$.  The black 
contour corresponds to $\Omega h^2 = 0.11$, the observed DM relic abundance.}}
\label{fig:Sigma}
\end{figure}

\begin{figure}[t]
\centering
\begin{tabular}{ccc}
\includegraphics[scale=0.65]{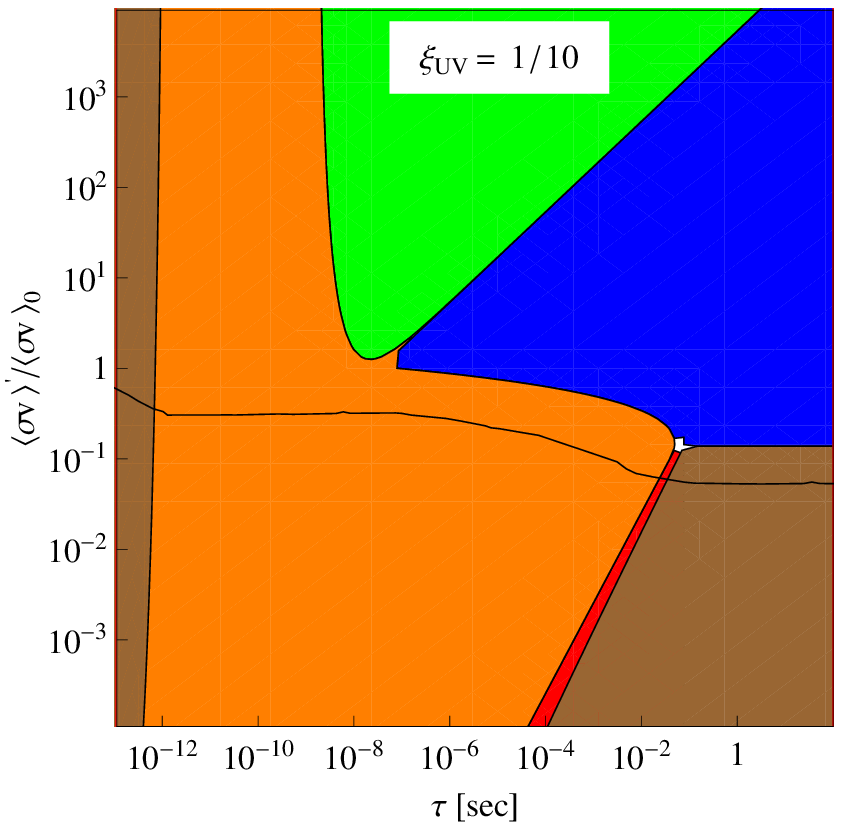}& 
\includegraphics[scale=0.65]{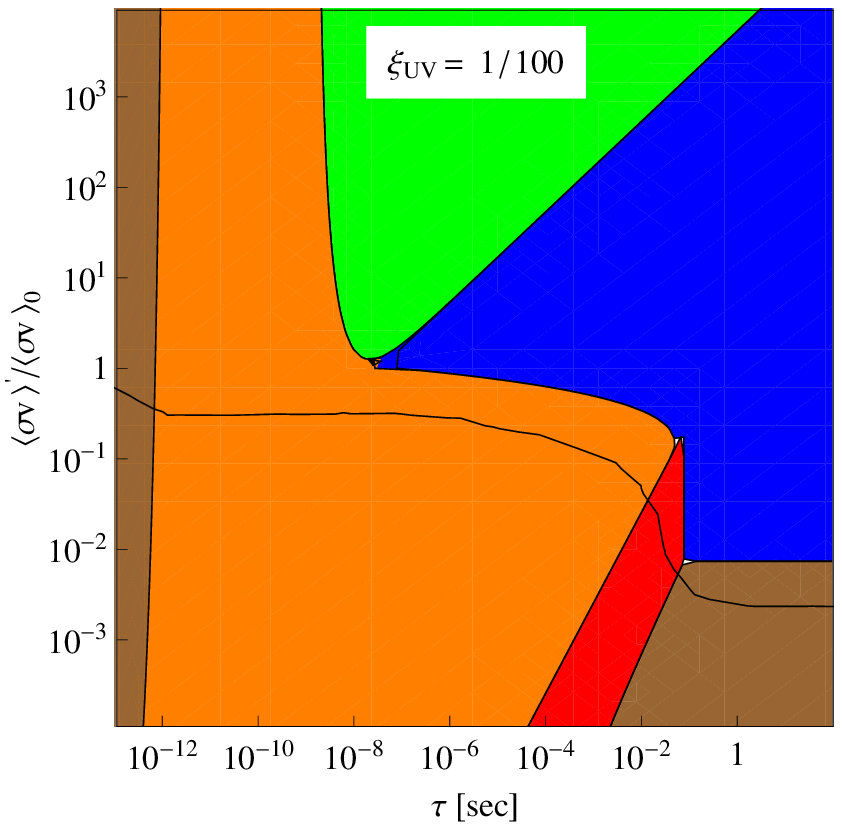}& 
\includegraphics[scale=0.65]{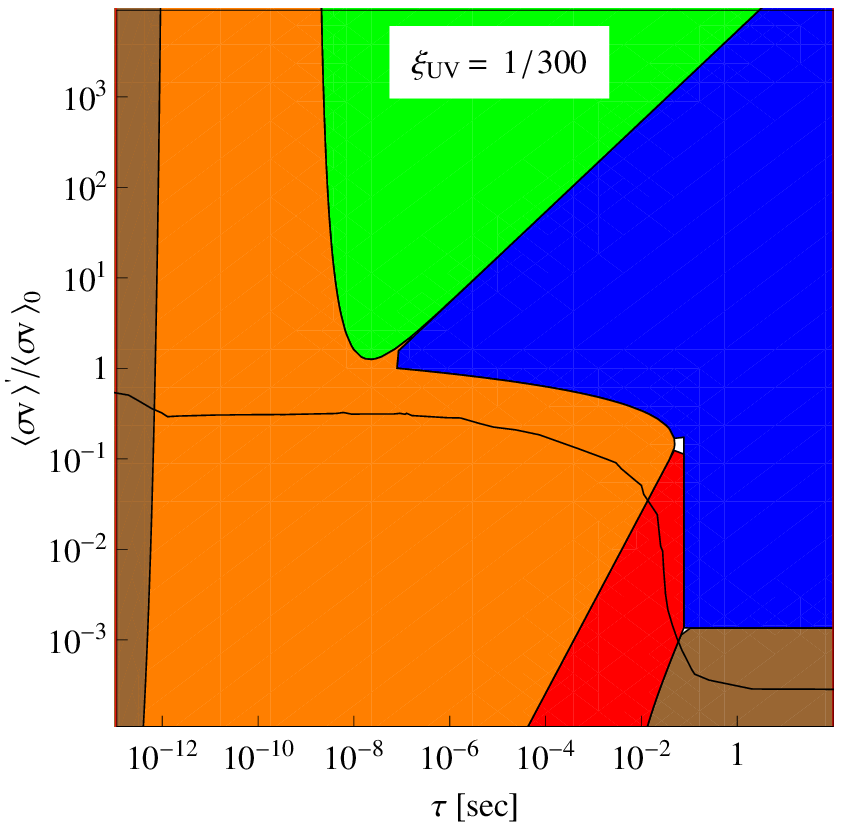} \\
\end{tabular}
\caption{\footnotesize{Sequence of cosmological phase diagrams with $\xi_{\rm UV}$ varying.  Other parameters are as in Figure \ref{phasediag}:  $(m=100\,\mathrm{GeV},\,m'=50\,\mathrm{GeV},\,\sv=3\times 10^{-26} \textrm{ cm}^3/\textrm{s})$.  The black 
contour corresponds to $\Omega h^2 = 0.11$, the observed DM relic abundance.}}
\label{fig:xi}
\end{figure}

\begin{figure}
\centering
\begin{tabular}{ccc}
\includegraphics[scale=0.65]{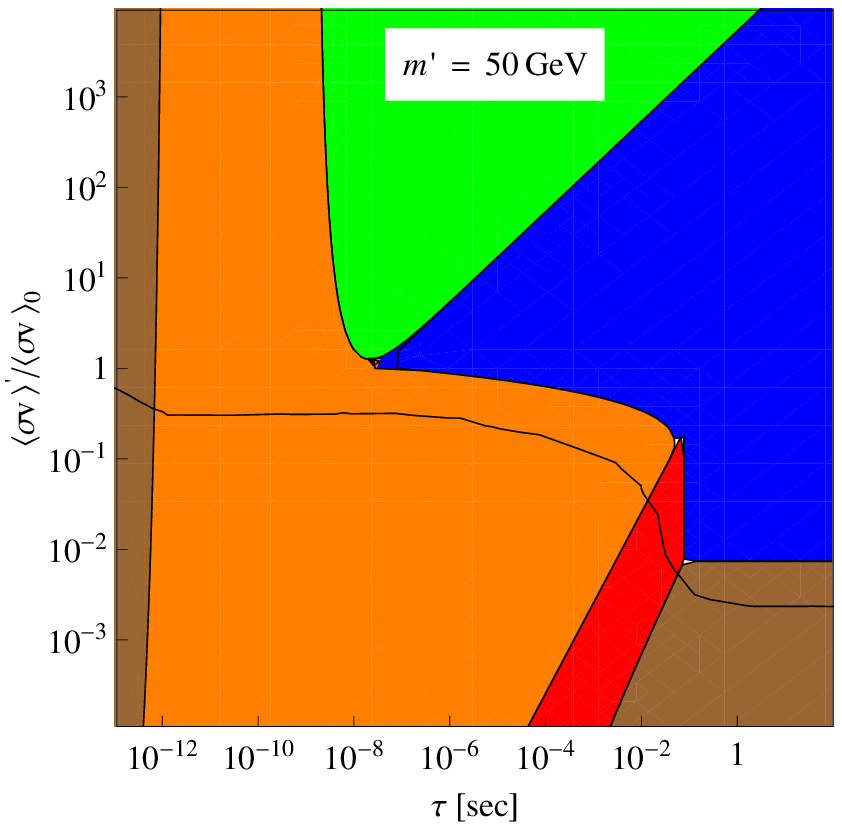}&
\includegraphics[scale=0.65]{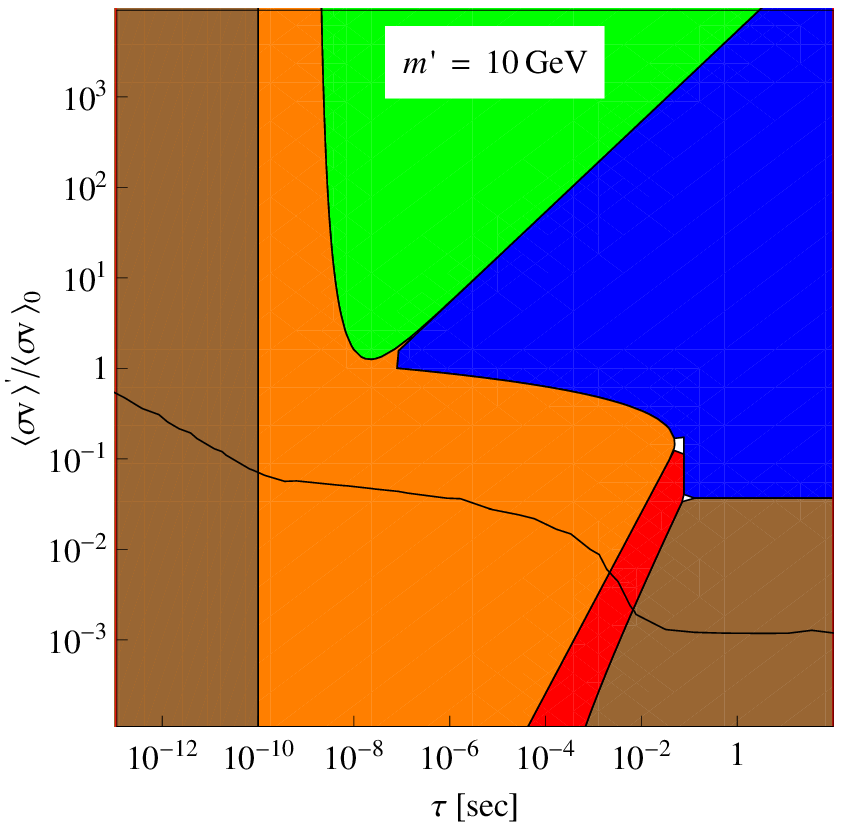} & 
\includegraphics[scale=0.65]{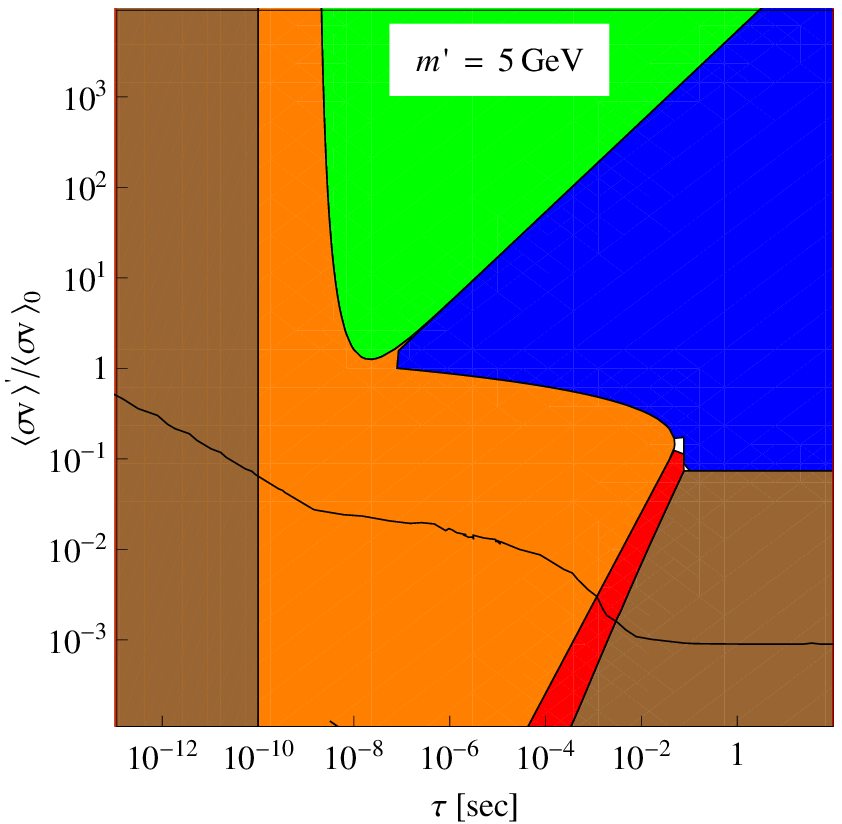} \\ 
\end{tabular}
\caption{\footnotesize{Sequence of cosmological phase diagrams with $m'$ varying.  Other parameters are as in Figure \ref{phasediag}:  $(\xi_{\rm UV}=0.01,\,m=100\,\mathrm{GeV},\,\sv=3\times 10^{-26} \textrm{ cm}^3/\textrm{s})$.  The black 
contour corresponds to $\Omega h^2 = 0.11$, the observed DM relic abundance.}}
\label{fig:Rm}
\end{figure}

In the cosmological phase space diagram depicted in Figure 7, the parameters $\{m, m^\prime, \sv, \xi_{\rm UV}\}$ were fixed. 
Let us now examine how each phase region changes as we vary these four parameters. 
Figures \ref{fig:Sigma}, \ref{fig:xi}, and \ref{fig:Rm} are cosmological phase diagrams showing the variation of one of these parameters while keeping the remaining
three fixed. The effects on the regions can be understood by referring to the formulae in Section \ref{cpd} which define the boundaries between regions. 
In addition, we would like to examine how the mechanism of 
DM production is affected.  To this end we include in these figures a single 
``critical DM abundance'' contour (black line) corresponding to the $\Omega h^2 = 0.11$.

Figure \ref{fig:Sigma} illustrates the effect of varying $\sv$. 
The most dramatic change is in the contours of total yield. 
As can be seen from the analytic yield formulas in Section \ref{cpd}, when $\sv$ is decreased,  
$\frac{\sv^{\prime}}{\sv}$ must increase in order to maintain a constant yield, whether dominated by 
FO$^\prime$, FO$\&$D, FO$\&$D$_{\rm r}$, or FI$_{\rm r}$. In addition the boundaries of the FI region shift to 
smaller lifetimes. Thus the yield contours ``rise'' in the $ \tau - \frac{\sv^{\prime}}{\sv}$ plane.   
For larger values of $\sv$, as in the left panel of Figure \ref{fig:Sigma}, the black line corresponding to $\Omega h^2 = 0.11$ can access the FI region but not the FO\&D region. Once FI dominates it will give the correct DM abundance when:
\be 
\tau &\simeq& (3\times 10^{-15} \text{ s})\times  \MPl \frac{m^\prime}{m^2} \frac{g}{g_{\star}^{3/2}} \\
&\simeq& \left( 4 \times 10^{-2} \text{ s} \right) \left(\frac{m'}{m} \right)\left( \frac{100 \text{GeV}}{m} \right)
 \left( \frac{228.5}{g_{\star}} \right)^{3/2}
 \label{eq:FIdom}
\ee
this corresponds to a lifetime of $\simeq 10^{-2}$ s in the left panel of Figure \ref{fig:Sigma}. 
As $\sv$ is decreased the critical DM contour rises and can start to access the FO$\&$D region. 
For FO$\&$D to give the correct DM abundance the following relation must 
hold:
\be
\frac{\sv m}{m^{\prime}} \sim \frac{4\times 10^{10}}{\MPl \sqrt{g_{\star}}} 
\sim \frac{2 \times 10^{-25} \, \text{cm}^3 \text{sec}^{-1}}{\sqrt{g_\star}}
\label{eq:FOdom}
\ee
This is the case for the center panel of Figure \ref{fig:Sigma} with $g_{\star} = 228.5$.
Decreasing $\sv$ even further leads to FO$\&$D$_{\rm r}$ domination, as shown in the right panel 
of Figure \ref{fig:Sigma}. 

Figure \ref{fig:xi} illustrates the effect of varying $\xi_{\rm UV}$. As is expected from Eq.(46)
decreasing $\xi_{\rm UV}$ results in a smaller large-$\tau$ $\FOp$ region. This makes sense since 
decreasing $\xi_{\rm UV}$ corresponds to an earlier $\FOp$  
(for large $\tau$, $\xi_{\rm FO^{\prime}} \sim \xi_{\rm UV}$)
and a smaller  $Y_{\FOp}'$ which will thus dominate in a smaller region of parameter space. It is clear that the FI region expands as $\xi_{\rm UV}$ is decreased.



Figure \ref{fig:Rm} illustrates the effect of varying $m^\prime$. 
From (\ref{eq:FIdom}) it is clear that the lifetime 
that yields the observed DM abundance from FI is proportional to $m^{\prime}$. 
In the left panel of Figure \ref{fig:Rm}, $m^{\prime} = 50$ GeV and the FI lifetime 
is $\tau \sim 10^{-2}$ s; decreasing $m^{\prime}$ it is possible to lower this lifetime. 
This is shown in the center and right panels of Figure \ref{fig:Rm}; $m^{\prime} = 10$ GeV 
requires $\tau \sim 10^{-3}$ s for FI to dominate. Also note that decreasing $m^{\prime}$ 
results in a larger FO$^{\prime}$ region at small $\tau$, as expected since this boundary
is approximately the line $\xi = \frac{m^{\prime}}{m}$.

Finally we comment on the variation of $m$. The only major effect is the change in the lifetime 
at the thermalization bound (red line) which corresponds to $\xi \left(\tau, m \right) = 1$. 
Thus as $m$ is increased this lifetime is decreased.

\section{Dark Matter Asymmetry}
\label{sec:dma}

Until now our discussion has been limited to cases in which the final abundance of the DM, $X'$, is symmetric under the exchange of particles and anti-particles.   As such we have implicitly assumed that either $X'$ is its own anti-particle or that the mechanism of DM production dynamically produces equal numbers of $X'$ and $\bar{X}'$.  In this section we determine the conditions necessary for DM to arise from  a particle anti-particle asymmetry
\be
\eta' = \frac{n' - \bar{n}'}{s}
\ee
rather than from a symmetric yield, $Y' = (n' + \bar{n}')/s$.  An asymmetric mechanism of DM production requires that the hidden sector possess a global $U(1)$ symmetry, $Q'$.  A crucial question is then whether the portal interactions connecting each sector either completely break $Q'$, or whether they preserve some combination of $Q'$, baryon number $B$, and lepton number $L$.  We study two possible scenarios in which there exists a global $U(1)$ symmetry $S$ having the properties
\begin{itemize}
\item $S_1=Q'$ rotates only hidden sector fields and is broken by the connector interactions.
\item $S_2= \alpha B + \beta L + \gamma Q'$, with real parameters $\alpha, \beta$ and $\gamma$, is exact, except possibly for anomalies.
\end{itemize}   

\subsection{Asymmetric FI and Asymmetric FO\&D}

At temperatures well above the weak scale we assume that there does not exist any asymmetry between the DM particles and anti-particles, so $\eta' = 0$.  Is it possible for a non-zero asymmetry to be generated by the thermal production mechanisms studied in the last section?  At first sight the answer is no, since each sector is separately in thermal equilibrium.  While thermal equilibrium is lost right at the end of FO and FO$'$, these processes involve {\it total} annihilation cross-sections $\sv, \sv'$ which are the same for anti-particles as for  particles, so an asymmetry cannot be generated.  However, the sectors are at different temperatures, so processes mediated by the connector interactions are not in thermal equilibrium.  This offers new possibilities for generating a DM asymmetry, and indeed for baryogenesis.  For the symmetry $S_1$, the connector interactions may generate $\eta'$, but it is unrelated to the baryon and lepton asymmetries, $\eta_{L,B}$.  On the other hand, symmetry $S_2$ is exact so any $Q'$ charge that is generated must be compensated by baryon and lepton charges, so that the universe remains neutral under the charge of $S_2$. Asymmetric FI that provides such a correlation between the baryon and DM densities is discussed in \cite{TBA}, and explicit supersymmetric models are constructed.

There are now four Boltzmann equations to be studied for the evolution of $n, \bar{n}, n'$ and $\bar{n}'$.  The equations for $n$ and $\bar{n}$ are identical and take the form of Eq.~(\ref{eq:Bn}).  The decay rate $\Gamma$ is the total decay rate of $X$, and by CPT the total decay rate of $\bar{X}$ is also $\Gamma$.  However, the equations for $n'$ and $\bar{n}'$ become more complicated once we consider multiple decay modes for $X$, which is one of the necessary conditions to generate an asymmetry.  We index each decay mode by $i$, with a  partial width $\Gamma_i$ corresponding to
\be
X \rightarrow p_i X' + \bar p_i \bar{X}' + (\mbox{particles with} \,\,\, Q'=0)
\label{eq:modei}
\ee
where $p_i$ and $\bar p_i$ are integers denoting the number of $X'$ and $\bar X'$ particles produced via the $i$th decay mode. The $X'$ and $\bar{X}'$ may result from a cascade of decays in the hidden sector.  
In general, there can be modes containing particles other than $X'$ and $\bar{X}'$ (say $Y'$) with $Q' \neq 0$; however we assume that the hidden sector interactions are large enough that $Y'$ rapidly decays into $X'$ and $\bar{X}'$, so that we only need to consider modes of the form in (\ref{eq:modei}) above.  If $X$ is not equal to its own anti-particle then $\bar{X}$ has a set of decay modes to the corresponding anti-particles with partial widths $\bar{\Gamma}_i$, and by CPT the total decay rates of $X$ and $\bar{X}$ are equal, i.e. defining $\Gamma \equiv \sum_i \Gamma_i$ and  $\bar \Gamma \equiv \sum_i  \bar{\Gamma}_i$, then $\Gamma = \bar\Gamma$.  On the other hand, if $X$ is real it can decay both via Eq.~(\ref{eq:modei}) and by the decay to the corresponding anti-particles, so that its total lifetime is $\Gamma = \sum_i \Gamma_i + \sum_i  \bar{\Gamma}_i$. 

The Boltzmann equations for $n'$ and $\bar{n}'$ are:
\be
\frac{d}{dt} n' + 3 H n' &=& - (n' \bar{n}'  - n'_{\rm eq} \bar{n}'_{\rm eq}) \sv' + n \sum_i p_i \Gamma_i  +  \bar{n} \sum_i \bar p_i \bar{\Gamma}_i  
\label{eq:ABn'}
\ee
\be
\frac{d}{dt} \bar{n}' + 3 H \bar{n}' &=& -(n' \bar{n}'  - n'_{\rm eq} \bar{n}'_{\rm eq}) \sv'  + n \sum_i \bar p_i \Gamma_i +  \bar{n} \sum_i p_i \bar{\Gamma}_i. 
\label{eq:ABn'bar}
\ee
As with the analysis of the symmetric abundance $Y'$ in the previous section, we omit scattering process contributions to FI. They are UV dominated both for $Y'$ and $\eta'$, and it is unclear how they can be distinguished experimentally from a high scale initial condition on $\eta'$.  
Taking sums and differences of these two equations, the source terms for $n' \pm \bar{n}'$ are $\sum_i(p_i \pm \bar p_i)(n \Gamma_i \pm \bar{n} \bar{\Gamma}_i)$. The goal of this analysis is to understand how $\eta'$ is generated from these equations starting from \emph{symmetric} boundary conditions, i.e. when $\bar{n}=n$. As such, this mechanism is markedly different from using these equations to transfer a pre-existing asymmetry in $X$ to $X'$, as has been studied in the context of ADM \cite{Kaplan:2009ag}.  With $\bar{n}=n$ the source terms become $\Gamma n \sum_i (p_i \pm \bar p_i)(r_i \pm \bar{r}_i)$, where we introduced the branching ratios $r_i = \Gamma_i/\Gamma$ and  $\bar{r}_i = \bar{\Gamma}_i/\bar{\Gamma}$. This last form is also the correct source term for the case that $X$ is the same as its anti-particle.

Using these source terms, the symmetric and asymmetric contributions to the DM yields are given for FO\&D by
\be
Y'_\textrm{FO\&D}  =  \frac{C_{\rm FO}}{\MPl T_\FO \sv } \sum_i(p_i + \bar p_i) (r_i + \bar{r}_i) \qquad  \eta'_\textrm{FO\&D}  =  \frac{C_{\rm FO}}{\MPl T_\FO \sv } \sum_i (p_i - \bar p_i)(r_i - \bar{r}_i),
\label{eq:AFOD}
\ee
and for FI by
\be
Y'_\FI  = C_{\rm FI} \frac{\MPl \Gamma}{m^2} \sum_i(p_i + \bar p_i) (r_i + \bar{r}_i)   \qquad \eta'_{FI}  = C_{\rm FI}  \frac{\MPl \, \Gamma}{m^2} \sum_i (p_i - \bar p_i)(r_i - \bar{r}_i).
\label{eq:AFI}
\ee
The symmetric yields $Y'_{\rm FO\&D}$ and $Y'_{\rm FI}$ were studied in great detail in Sections \ref{sec:twosec} and \ref{cpd}. Note that these yields do not take re-annihilation into account; this will be discussed soon. On the other hand, the asymmetric yields $\eta'_{\rm FO\&D}$ and $\eta'_{\rm FI}$, are new---we denote these DM production mechanisms by Asymmetric Freeze-Out and Decay (FO\&D$_{\rm a}$) and Asymmetric Freeze-In (FI$_{\rm a}$).  
As seen from (\ref{eq:AFOD}) and (\ref{eq:AFI}), generating an asymmetry requires decay modes with $\bar p_i \neq p_i$ so that the final state has $Q' \neq 0$, as well as $\bar{r}_i \neq r_i$.

According to Eqs.~(\ref{eq:AFOD}) and (\ref{eq:AFI}), the symmetric and asymmetric mechanisms obey the general relation,
\be
\eta' & =&  \epsilon \, Y'  ,
\label{eq:eta'}
\ee
where 
\be
\epsilon \, = \, \sum_i \frac{(p_i - \bar p_i)(r_i - \bar{r}_i)}{(p_i + \bar p_i) (r_i + \bar{r}_i)}.
\label{eq:eps}
\ee
Here $\epsilon$ is a general measure of CP violation occurring in decays of $X$. 
If, for example, there are two relevant decay modes such that $p_{1,2}$ and $\bar p_{1,2}$ are not large, then $\epsilon$  is roughly given by  $r_1 - \bar{r}_1$.  

Next, let us determine the typical size of $\epsilon$ for a simple case with two decay modes.  We take the two decay modes of interest to be $X \rightarrow X' + f_1$ and $X \rightarrow Y' + f_2$, where $f_i$ are particles in the final state that have $Q'=0$, and $Y'$ and $X'$ have different $Q'$ charge.  We introduce dimensionless amplitudes $A_i$ to describe these decays, defined by $\Gamma_i = |A_i|^2 m/8 \pi$. Furthermore, as mentioned earlier we assume the existence of additional (hidden) interactions which allow for the rapid decay of $Y'$ via $Y' \rightarrow p_2 X' + \bar p_2 \bar{X}'$, with $p_2-\bar{p}_2 \neq 1$. In addition, a vertex allowing for the rescattering process $Y' + f_2 \rightarrow X' + f_1$ (with dimensionless amplitude $A_{12}$) is also required.  This allows the final state of process 2 above to rescatter into the final state of process 1 at one loop (and vice versa), which is necessary for successful asymmetry production.

If the aforementioned amplitudes are too large, they will cause the two sectors to equilibrate at the era $T \approx m$ and destroy the viability of the FI and FO\&D mechanisms.  Applying the equilibration condition Eq.~(\ref{eq:taumin}) demands that
\be
|A_i| \lesssim 10^{-6} \sqrt{\frac{m}{100 \,\mbox{GeV}}}  \sqrt{\frac{g'_*(T \simeq m)/g_X}{100} }.
\label{eq:Ai}
\ee
A non-zero value for $\epsilon$ results from interference between tree and one loop contributions to the decays
\be
\epsilon \simeq \frac{1}{16 \pi} \; \frac{{\rm Im} (A_1 A_2^* A_{12})}{|A_1|^2 + |A_2|^2}.
\ee
In general the rescattering involves both visible and hidden sector particles so that, to avoid equilibration of the sectors at $T\approx m$, $A_{12}$ must satisfy the same bound, (\ref{eq:Ai}), as $A_i$, giving
\be
\epsilon \lesssim 10^{-8} \sin \phi \; \sqrt{\frac{m}{100 \,\mbox{GeV}}}  \sqrt{\frac{g'_*(T \simeq m)/g_X}{100} }
\label{eq:eptherm}
\ee
where $\phi = \arg(A_1 A_2^* A_{12})$.  

Is it possible to evade this bound? For theories with a global symmetry of type $S_2$ such that a combination of $B$, $L$ and $Q'$ is preserved, the requirement that the two decay modes have different $Q'$ charge implies that the two modes also have different $B/L$ charge so that the set of particles comprising $f_1$ and $f_2$ are different. This further implies that rescattering $A_{12}$ between the final states of the two decay modes involves \emph{both} visible and hidden sector particles so that it is not possible to evade the bound (\ref{eq:eptherm}) above. 

However, this bound may be evaded in theories where the global symmetry is of type $S_1$ since then it is possible for $f_1$ and $f_2$ to contain the same set of visible sector particles. Rescattering then only involves the hidden sector and $A_{12}$ can be ${\cal O}(1)$ in principle.  There is still the requirement that the rescattering amplitude not wash-out the asymmetry once it is produced, but this is highly model dependent since the asymmetry may be produced at $T' \ll m_{X'}$, so that the washout is exponentially suppressed.  Hence, in these theories $\epsilon$ can be as large as $10^{-2}$ for $A_{12}$ of ${\cal O}(1)$.

Can these new mechanisms generate sufficient DM?  This requires $m'\eta' =4\times  10^{-10}\textrm{ GeV}$.   For Asymmetric FI
\be
m' \eta'_{\rm FI} \simeq 4\times 10^{-10}\textrm{ GeV}\; \left(\frac{10^{-10} \textrm{ s}}{\tau} \frac{|A_{12}|}{10^{-6}} \right) \left( \frac{C_{\rm FI}}{10^{-3}} \right)\left(\frac{m'}{40\textrm{ GeV}}\right)\left( \frac{100 \, \mbox{GeV}}{m} \right)^2 \frac{\sin \phi|A_1 A_2|}{|A_1|^2 + |A_2|^2}\nonumber
\ee
while for Asymmetric FO\&D
\be
m'\eta'_{\rm FO\&D} \simeq 4\times 10^{-10}\textrm{ GeV}\;\left(\frac{10^{-8} \sv_0}{\sv} \frac{|A_{12}|}{10^{-6}} \right)\left(\frac{C_{\rm FO}}{10^{-1}}\right) \left(\frac{m'}{40\textrm{ GeV}}\right) \left( \frac{100 \, \mbox{GeV}}{m} \right) \frac{\sin \phi|A_1 A_2|}{|A_1|^2 + |A_2|^2}. \nonumber
\ee
Since the last factor in these equations is always less than unity, Asymmetric FI requires a short lifetime $\tau< 10^{-10} \textrm{ s} \; |A_{12}|/ 10^{-6}$ and Asymmetric FO\&D requires a small annihilation cross-section $ \sv < 10^{-8} \sv_0 \; |A_{12}|/ 10^{-6}$. 

\begin{figure}[t]
\centering
\begin{tabular}{cc}
\includegraphics[scale=1]{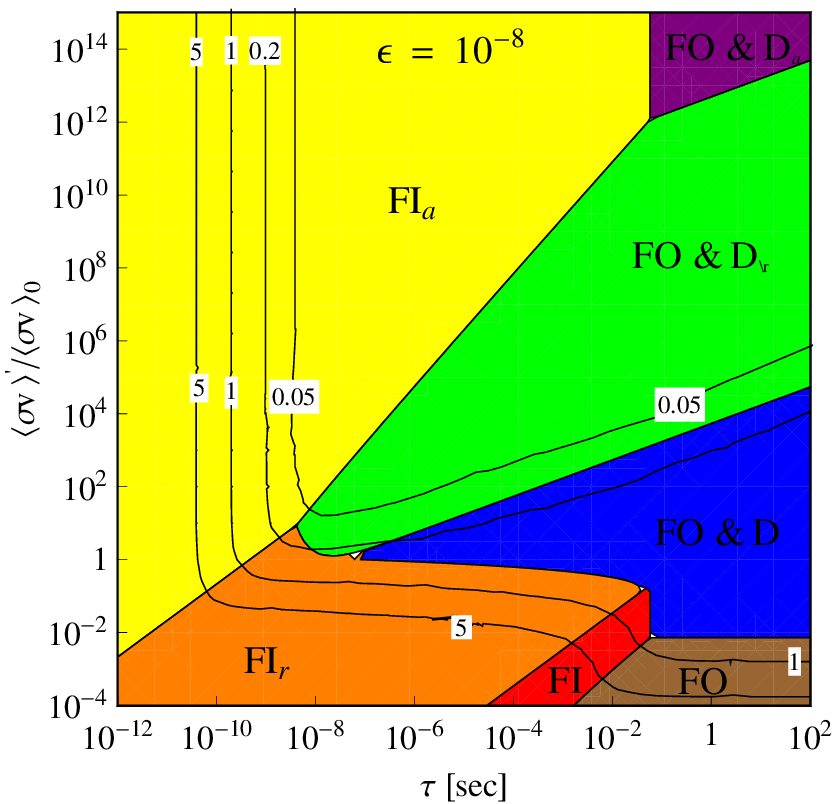}& 
\includegraphics[scale=1]{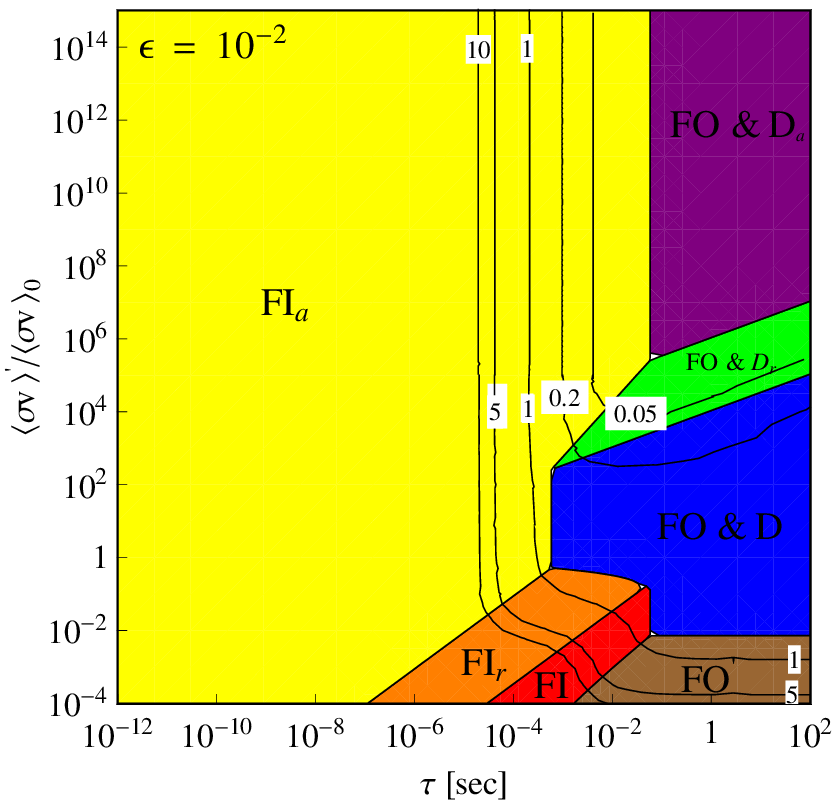}\\
\end{tabular}
\caption{\footnotesize{Cosmological phase diagrams similar to Figure \ref{phasediag}, only depicting FO\&D$_{\rm a}$ and FI$_{\rm a}$ along with other production mechanisms.  Here the numerically evaluated contours correspond to $\Omega h^2$ arising from the sum of $Y'$ and $\eta'$.  The left and right panels correspond to $\epsilon=10^{-8}\textrm{ and } 10^{-2}$, respectively.  }}
\label{fig:phasediag2}
\end{figure}

The symmetric yields $Y'$ in (\ref{eq:AFOD}) and (\ref{eq:AFI}) do not take re-annihilation into account. As has been discussed earlier in section \ref{sec:reann}, if the (symmetric) yields for FO\&D and $\FI$ are sufficiently large, then re-annihilation occurs giving rise to much smaller values for the final (symmetric) yields
$Y'_{\mathrm{FO\&D_r}}$ and $Y'_{\FI_r}$. This is crucial for the asymmetric yield to dominate the symmetric yield since $\epsilon$ arises at the loop level and is expected to be small. DM will be dominated by the asymmetric component of $X'$ only if $\sv'$ is sufficiently large for re-annihilation to occur, hence reducing the symmetric component $Y'$ while leaving the asymmetric component  $\eta'$ unaltered.  Thus a DM asymmetry can be the dominant component of DM only in the re-annihilation regions for the symmetric abundance depicted in Figure  \ref{phasediag}. It is interesting that while FI dominates the DM density only in a narrow region of parameters as shown in Figure \ref{phasediag}, asymmetric FI can dominate over a very wide region of parameters (assuming other conditions for generating the asymmetry are satisfied).   

\subsection{Cosmological Phase Diagrams for DM Asymmetries}

Cosmological phase diagrams are shown in Figure \ref{fig:phasediag2} for non-zero values of $\epsilon$.  The parameters held fixed are the same as in Figure \ref{phasediag}, allowing a comparison of the asymmetric and symmetric DM production mechanisms.
The contours are of $\Omega h^2$, now arising from the sum of $Y'$ and $\eta'$, and we have included {\it purple} and {\it yellow} regions denoting FO\&D$_{\rm a}$ and FI$_{\rm a}$ phases, respectively.  The left and right panels of Figure \ref{fig:phasediag2} correspond to $\epsilon = 10^{-8}, 10^{-2}$.  Here the lower value of $\epsilon$ arises from the non-thermalization bound in Eq. (\ref{eq:eptherm}) and the upper value arises in theories of type $S_1$ where $\epsilon$ is only bounded from above by the fact that in the perturbative regime, CP violating decays must always be at least a loop factor down from the CP respecting decays.

In Figure \ref{fig:phasediag2} we see that the regions corresponding to FO\&D and FI  are essentially unchanged from Figure \ref{phasediag}, while the regions for FO\&D$_{\rm a}$ and FI$_{\rm a}$ are shown encroaching on the FO\&D$_{\rm r}$ and FI$_{\rm r}$ regions.  This is reasonable because these asymmetric DM production mechanism are only active when re-annihilation effects are maximal, so the symmetric component of the DM relic abundance is efficiently destroyed.  As expected, for larger $\epsilon$, as depicted in the right panel of Figure \ref{fig:phasediag2}, FI$_{\rm a}$ becomes dominant at larger values of $\tau$, which is reasonable because in this case the CP phase in $X$ decays is larger so the decay rate of $X$ can be smaller while producing the same asymmetric DM abundance.  On the other hand, at large $\epsilon$, FO\&D$_{\rm a}$ can dominate at smaller values of $\sv ' / \sv$, since the large CP phase in $X$ decays allows for the abundance of $X$ arising from FO to be smaller while producing the same DM asymmetry from late decays.  There are no contours in the FO\&D$_{\rm a}$ region because this production mechanism gives an abundance independent of $\sv'$ and $\tau$.

Note the behavior of the contours of $\Omega h^2$ depicted in Figure \ref{fig:phasediag2}.  In particular, as soon as these curves enter the FI$_{\rm a}$ region, they turn vertically upward. This is an indication that the abundance of DM contributed by Asymmetric FI is entirely independent of $\sv '/ \sv$, and instead only depends on $\tau$---thus, FI$_{\rm a}$ behaves essentially identically to FI without re-annihilation effects. 

Next, let us consider the boundary between Asymmetric FI and Asymmetric FO\&D.  In particular, the abundance from FI$_{\rm a}$ dominates the contribution from FO\&D$_{\rm a}$ when  $\eta'_{\rm FI} >  \eta'_\textrm{FO\&D}$, so
\be
\tau \, < \, C_{\rm FI} \,\,  \frac{\MPl}{m^2 Y'_\textrm{FO\&D}} \, = \, \frac{C_{\rm FI}}{C_{\rm FO} x_{\rm FO}} \, \frac{\MPl^2 \sv}{m}.
\label{eq:APD}
\ee
Thus FO\&D$_{\rm a}$ dominates at large $\tau$, since then FI$_{\rm a}$ is negligible, and for $\sv \ll \sv_0$, so there is a very large FO abundance of $X$.  Notice that the right-hand side of Eq.~(\ref{eq:APD}) depends on quantites which are fixed between the two panels of Figure \ref{fig:phasediag2}, which is why the boundary between FI$_{\rm a}$ and FO\&D$_{\rm a}$ is at the same value of $\tau$ in both.

FO\&D$_{\rm a}$ requires very small $\sv$, especially if $\epsilon$ is small, and also dominates for large $\tau$ where it might lead to signatures in big bang nucleosynthesis.  On the other hand, FI$_{\rm a}$ is independent of $\sv$ and can successfully account for DM over a very wide range of $\tau$.  The challenge for both asymmetric mechanisms is to construct theories where the CP violating phase in $\epsilon$ can be measured in the laboratory.

\section{Conclusions}
\label{sec:conclusions}

The nature and origin of DM remains a deep mystery. A priori, there is no reason for DM to couple to the visible sector with interactions other than gravity, as only the gravitational properties of DM have been definitively verified through astrophysical probes.  As such, one seeks theoretical motivations for why DM should be anticipated at other laboratory experiments, for example in direct detection,  particle colliders, or indirect signals from cosmic rays. 

One popular theoretical justification is the so-called ``WIMP miracle'', whereby DM is initially in thermal equilibrium with visible sector particles and undergoes FO, yielding approximately the correct DM relic abundance for weak scale masses and cross-sections.  This setup has the advantage that DM generation is ``IR dominated'', and fixed wholly by the DM annihilation cross-section, which may be directly accessible at colliders.

That said, it is clear that the single sector setup required for the WIMP miracle is only a very 
particular slice within the totality of theory space.  In general, extra dimensional and string theoretic constructions motivate the existence of one or more hidden sectors which are very weakly coupled to our own visible sector.  The inclusion of a weak-scale hidden sector, complete with its own set of particles, dynamics, and thermal history, substantially expands the allowed space of mechanisms for DM generation.  Concretely, single sector FO is extended to two sector FO\&D and FI, along with their re-annihilated and asymmetric cousins.  

Despite the proliferation of DM production modes, this broad two sector framework retains the IR dominated features lauded in the WIMP miracle.  As we have shown, since both the visible and hidden sectors are initiated in a thermal state, the origin of DM and its final abundance are dictated entirely by the handful of quantities shown in Eq.~(\ref{eq:params}).  Remarkably, many of these quantities, namely $m$, $m'$, $\sv$, and $\tau$ may be measured at particle colliders!  This offers the exciting prospect that the origin of DM might be successfully reconstructed from collider physics in this enormous and theoretically motivated class of theories.  We consider the details of such an endeavor in a companion paper \cite{colliderpaper}.



 \section*{Acknowledgments}
L.H. thanks John March-Russell and Stephen West for useful discussions.
This work was supported in part by the Director, Office of Science, 
Office of High Energy and Nuclear Physics, of the US Department of 
Energy under Contract DE-AC02-05CH11231 and by the National Science 
Foundation on grant PHY-0457315. The work of PK is also supported in part by
the US Department of Energy contract DE-FG02-92ER-40699.


\begin{thebibliography}{99}

\bibitem{Kolb:1988aj}
  E.~W.~Kolb and M.~S.~Turner,
  ``The Early Universe.'' (1988).

\bibitem{Feng:2003uy}
  J.~L.~Feng, A.~Rajaraman and F.~Takayama,
  Phys.\ Rev.\  D {\bf 68}, 063504 (2003)
  [arXiv:hep-ph/0306024].

\bibitem{Feng:2004zu}
  J.~L.~Feng, S.~f.~Su and F.~Takayama,
  Phys.\ Rev.\  D {\bf 70}, 063514 (2004)
  [arXiv:hep-ph/0404198].

\bibitem{Hall:2009bx}
  L.~J.~Hall, K.~Jedamzik, J.~March-Russell and S.~M.~West,
  JHEP {\bf 1003}, 080 (2010)
  [arXiv:0911.1120 [hep-ph]].

\bibitem{Asaka:2005cn}
  T.~Asaka, K.~Ishiwata and T.~Moroi,
  Phys.\ Rev.\  D {\bf 73}, 051301 (2006)
  [arXiv:hep-ph/0512118].


\bibitem{Feng:2008mu}
  J.~L.~Feng, H.~Tu and H.~B.~Yu,
  JCAP {\bf 0810}, 043 (2008)
  [arXiv:0808.2318 [hep-ph]].

\bibitem{Kaplan:2009ag}
  D.~E.~Kaplan, M.~A.~Luty and K.~M.~Zurek,
  Phys.\ Rev.\  D {\bf 79}, 115016 (2009)
  [arXiv:0901.4117 [hep-ph]].
\\
  Y.~Cai, M.~A.~Luty and D.~E.~Kaplan,
  arXiv:0909.5499 [hep-ph].


\bibitem{colliderpaper}
  C.~Cheung, G.~Elor, L.~J.~Hall, P.~Kumar,
  JHEP {\bf 1103}, 085 (2011).
  [arXiv:1010.0024 [hep-ph]].


\bibitem{Amsler:2008zzb}
  C.~Amsler {\it et al.}  [Particle Data Group],
  Phys.\ Lett.\  B {\bf 667}, 1 (2008).

\bibitem{Asaka:2000ew}
  T.~Asaka and T.~Yanagida,
  Phys.\ Lett.\  B {\bf 494}, 297 (2000)
  [arXiv:hep-ph/0006211].

\bibitem{Cheung:2010mc}
  C.~Cheung, Y.~Nomura and J.~Thaler,
  JHEP {\bf 1003}, 073 (2010)
  [arXiv:1002.1967 [hep-ph]].

\bibitem{Cheung:2010qf}
  C.~Cheung, J.~Mardon, Y.~Nomura and J.~Thaler,
  JHEP {\bf 1007}, 035 (2010)
  [arXiv:1004.4637 [hep-ph]].

\bibitem{Moroi:1993mb}
  T.~Moroi, H.~Murayama and M.~Yamaguchi,
  Phys.\ Lett.\  B {\bf 303}, 289 (1993).

\bibitem{TBA}
  L.~J.~Hall, J.~March-Russell and S.~M.~West,
  arXiv:1010.0245 [hep-ph].



\end{thebibliography}
\end{document}